\documentclass[useAMS,usenatbib,twocolumn]{mn2e}
\pdfoutput=1
\usepackage{color,url,hyperref,epsfig,graphicx,amsmath,subfigure,amssymb}
\usepackage{morefloats,aas_macros,longtable}
\newcommand{\gae}{\lower 2pt \hbox{$\, \buildrel {\scriptstyle >}\over {\scriptstyle
\sim}\,$}}
\newcommand{\lae}{\lower 2pt \hbox{$\, \buildrel {\scriptstyle <}\over {\scriptstyle
\sim}\,$}}

\newcommand{\red}[1]{\textcolor{black}{#1}}

\newcommand{\logg}{log\,{\rm g}}
\newcommand{\teff}{\rm T_{eff}}
\newcommand{\logteff}{log\,{\rm T_{\rm eff}}}

\newcommand {\reac}[6] {$\rm\,{}^{#2}\kern-0.8pt{#1}\,({#3}\,,{#4})  \,{}^{#6}\kern-0.8pt{#5}\,$}

\newcommand{\Msun}{\mbox{${\rm~M_\odot}\,$}}

\newcommand{\Msolar}{\mbox{$M_{\odot}$}}

\newcommand{\sub}[1]{\mbox{$_{\rm #1}$}}

\newcommand{\Teff}{\mbox{$T\sub{eff}$}}

\newcommand{\logL}{\mbox{$\log(L/L_{\odot})$}}
\newcommand{\HP}{\mbox{$H_{\rm{P}}\,$}}

\newcommand{\beq}{\begin{equation}}
\newcommand{\eeq}{\end{equation}}
\newcommand{\beqa}{\begin{eqnarray}}
\newcommand{\eeqa}{\end{eqnarray}}
\newcommand{\benu}{\begin{enumerate}}
\newcommand{\eenu}{\end{enumerate}}
\newcommand{\bite}{\begin{itemize}}
\newcommand{\eite}{\end{itemize}}
\newcommand{\bdes}{\begin{description}}
\newcommand{\edes}{\end{description}}

\newcommand{\comment}[1]{}
\newcommand{\Mini}{$M_{\rm ini}$}

\title[PARSEC evolutionary tracks of massive stars]{{\sl PARSEC} evolutionary tracks of massive stars up to 350\,$\bmath{\Msolar}$ at metallicities 0.0001$\bmath{ \leq Z \leq}$0.04}

\author[Y. Chen et al.]{Yang Chen,$^{1,2}$\thanks{E-mail: ychen@sissa.it (CY); CY is supported by the SISSA-USTC joint doctoral program.}
  Alessandro Bressan,$^{2}$ L{\'e}o Girardi,$^3$ Paola Marigo,$^4$ Xu Kong,$^{1,5}$ \and Antonio Lanza$^{2}$ \\
  $^{1}$ Department of Astronomy, University of Science and Technology of China (USTC), Hefei 230026, Anhui, China\\
  $^{2}$ SISSA, via Bonomea 265, I-34136 Trieste, Italy\\
  $^{3}$ Osservatorio Astronomico di Padova -- INAF, Vicolo dell'Osservatorio 5, I-35122 Padova, Italy \\
  $^{4}$ Dipartimento di Fisica e Astronomia, Universit\`a di Padova, Vicolo dell'Osservatorio 2, I-35122 Padova, Italy\\
  $^{5}$ Key Laboratory for Research in Galaxies and Cosmology, USTC, Chinese Academy of Sciences, Hefei 230026, Anhui, China
}

\begin{document}
\date{}
\pagerange{\pageref{firstpage}--\pageref{lastpage}} \pubyear{2015}

\maketitle

\label{firstpage}

\begin{abstract}
We complement the PARSEC data base of stellar evolutionary tracks
with new models of massive stars, from the pre-main
sequence phase to the central Carbon ignition.
We consider a broad range of metallicities, 0.0001$\leq Z \leq$0.04 and initial masses up to
\Mini=$350\,\Msolar$.  The main difference with respect to our previous
models of massive stars is the adoption of a recent  formalism accounting for
the mass-loss enhancement
when the ratio of the stellar to the Eddington luminosity, $\Gamma_e$, approaches unity.
With this new formalism, the models are able to reproduce the Humphreys-Davidson
limit observed in the Galactic and Large Magellanic Cloud colour-magnitude diagrams,
without an {\sl ad hoc} mass-loss enhancement.
We also follow the predictions of recent wind models indicating that the
metallicity dependence of the mass-loss rates
becomes shallower when  $\Gamma_e$ approaches unity.
We thus find that the more massive stars may suffer from substantial mass-loss even at low metallicity.
We also predict that the Humphreys-Davidson limit
should become brighter at decreasing metallicity.
We supplement the evolutionary tracks with new tables of theoretical bolometric
corrections, useful to compare tracks and isochrones with the
observations.  For this purpose, we homogenize existing stellar
atmosphere libraries of hot and cool stars (PoWR, ATLAS9 and Phoenix)
and we add, where needed, new atmosphere models computed with
{\sl WM-basic}.  The mass, age and metallicity grids are fully adequate to
perform detailed investigations of the properties of very young
stellar systems, both in local and distant galaxies.
The new tracks supersede the previous old Padova models of massive stars.
\end{abstract}


\clearpage

\section{Introduction}

Massive stars play an important role in the evolution of galaxies.
They are the most important stellar sources of ionizing and
dissociation photons~\citep{Schaerer2011, Kimm2014, Dale2012, Cai2014,
  PDR1999, Yu2015}. They inject a significant amount of kinetic energy
through powerful stellar winds~\citep{Mackey2014}.  They are among the
main drivers of metal and dust enrichment in galaxies when they
explode as core collapsed supernovae (SNe)~\citep{Sarangi2015,
  Schneider2004}. They are thus very important sources of feedback to
the ambient ISM~\citep{Dale2013}.  Last but not least, most of our
information on the ongoing star formation rates across the Universe heavily
relies on our detailed knowledge of their properties~\citep{BC03, Kennicutt1998}.

Because of their relevance for so many fields of astrophysics, massive
stars have been the subject of many observational and theoretical
investigations that are impossible to list here.  Understanding their
evolution is challenged by the complexity of several physical
phenomena, as recently reviewed, for example, by
\citet{Martins&Palacios2013} who compare STERN~\citep{Brott2011},
Geneva~\citep{Ekstrom2012}, FRANEC~\citep{Chieffi_etal2013},
Padova~\citep{Bertelli2009}, MESA~\citep{xMESA2011} and
STAREVOL~\citep{xSTAREVOL2009} evolutionary tracks of massive stars.
\citet{Martins&Palacios2013} find that, apart from the inclusion of rotation,
the main differences among the
models computed by different authors can be attributed to the different
treatments of convection and mass-loss, with the former process being
more important in the domain of the less massive stars.
%
%
It is worth
noting that the Padova tracks~\citep{Bertelli2009} analyzed by
\citet{Martins&Palacios2013} are still those
presented in ~\citet{Fagotto_etal94} and~\citet{Bressan1993}.
Since then, there have been many advances both in the basic input
physics (opacities, equation of state, nuclear reactions) and in the
mass-loss theory (and observations), and these called for a substantial
revision of the Padova code.
But, while low and intermediate mass stars were systematically
updated, with the last version being  PARSEC\,({\sl PA}dova {T\!\sl\,R}ieste
{\sl\,S}tellar {\sl\,E}volution {\sl\,C}ode) models \citep[][and
  refs. therein]{PARSEC}, massive stars have not been updated since then.

In this paper we present the new evolutionary tracks of
massive stars computed with PARSEC. Besides
the basic input physics, which is described elsewhere \citep[][and
  refs. therein]{PARSEC}, the main novelties concern
the recipes adopted for the mass-loss rates
and  the spectral energy distributions used
to convert the tracks from the theoretical to the observational plane.
Concerning mass-loss, there have been many efforts over the past years
to determine/predict the mass-loss rates of massive stars across
different spectral types and metallicities.  It is now widely accepted
that hot massive blue supergiants (BSG) and Wolf-Rayet (WR) stars
lose a prominent amount of their mass through line-driven stellar
winds and that the mass-loss rates show a simple scaling law with the
metallicity ~\citep{CAK, Kudritzki2000, Nugis2000, Vink2000, Vink2001,
  Muijres2012b, WRreview, Smith2014}.  In Luminous Blue Variable (LBV)
stars, there is evidence that an important mode of mass-loss is
through eruptive mass-loss, that may contribute as much as or even
more than the steady stellar wind~\citep{Smith2009}.  In this phase
the mass-loss rate may easily reach a few of $10^{-4} \Msolar\,
yr^{-1}$~\citep{Lamers_LBV_mdot1989}.  This eruptive mechanism is
still unknown but the observed rates could be explained by a
super-Eddington wind~\citep{Smith_Owocki2006}, or by non-disruptive
hydrodynamic explosions~\citep{LBV_explostion_obs2014}. Interestingly,
LBV stars are found near the so-called Humphreys-Davidson
limit~\citep{Humphreys_Davidson_1979} which delimits the forbidden region above which only very few
stars are observed in the Hertzsprung-Russell (HR) diagram
of the Galactic massive stars.
In this respect, a remarkable result of recent investigations is that
the mass-loss rates could be enhanced by a significant factor when the
stars approach the Eddington luminosity \citep{Grafener2008,
  Vink2011}, which is known to happen near the
\citet{Humphreys_Davidson_1979} limit.  For the later spectral types
there are larger uncertainties both on the mechanisms and on the
strength of the mass-loss rates.  In red supergiants (RSG) one
customarily adopts the observational parametrization by
\citet{deJager1988}, but the mass-loss rates in this phase are known
to be uncertain by a large factor \citep{Salasnich1999, Meynet2015}. As
in less massive Asymptotic Giant Branch (AGB) stars, dust formation on the
circumstellar envelopes could be one of the possible mechanisms
responsible for this enhancement \citep{vanLoon2005}.

Concerning the atmosphere models used to predict the stellar magnitudes and colours,
they are equally important for the interpretation of observed properties of
massive stars.  For stars with negligible mass-loss, a comprehensive stellar
atmosphere library usually adopted is ATLAS9
\citep{Castelli2004}, consisting of plane parallel models in local
thermodynamic equilibrium.  This library is particularly suitable for
$\sim$ A, F and G type stars.
For cool giants, where the plane parallel and
non-LTE (non-local thermodynamic equilibrium) approximations must be relaxed,
comprehensive stellar atmosphere libraries are provided by
Phoenix \citep[][and references therein]{Phoenix} and MARCS \citep{marcs2008} projects.
For hot stars with high mass-loss rates a number of atmosphere models have been released in the recent
years like the {\sl WM-basic} models~\citep{WMbasic}, the CMFGEN models
\citep{CMFGEN} and the Potsdam Wolf-Rayet (PoWR) models
\red{\citep{PoWR1,PoWR2,PoWR3,PoWR4,PoWR5,PoWR6}}.

The paper is organized as following. In section 2, we describe our
stellar evolution models for massive stars, with particular care to
the description of the recipes adopted for the mass-loss rates.
In this section we also compare the new models with our previous
models and with  one set among the most recent models found in literature,
FRANEC~\citep{Chieffi_etal2013}.  In Section 3 we
describe in detail the adopted atmosphere models and the procedure used
to obtain as much as possible homogeneous sets of bolometric correction tables.
Finally, in section 4, we discuss the resulting colour-magnitude
diagrams predicted from the new evolutionary tracks and isochrones.

Our evolutionary tracks can be downloaded from
\url{http://people.sissa.it/~sbressan/parsec.html}, and the isochrones
can be downloaded from \url{http://stev.oapd.inaf.it/cgi-bin/cmd}.

\section{Stellar evolutionary tracks}
\label{sec:2}
We compute new tracks of massive stars with the PARSEC code
for a wide range of initial metallicities and for initial masses
from \Mini=14\,M$_\odot$ to \Mini=350\,M$_\odot$.
The evolution begins from the
pre-main sequence phase and ends at central Carbon ignition.
The new evolutionary tracks complement the already existing
models of intermediate and low mass stars.

\subsection{Basic input physics}
The input physics used in PARSEC is already thoroughly described
in \citet{PARSEC}, \citet{Bressan_etal13} and~\citet{Chen2014}.  Below
we briefly summarize the main points with particular attention to
those more relevant to the massive stars.  The equation of state (EOS)
is computed with the FreeEOS code of
A.W.~Irwin\footnote{http://freeeos.sourceforge.net.}.  Radiative opacities are
from the OPAL project \citep{IglesiasRogers_96} and, in the
low-temperature regime, from the \AE
SOPUS\footnote{http://stev.oapd.inaf.it/aesopus.} project
\citep{MarigoAringer_09}.  Conductive opacities are included
following \cite{Itoh_etal08}.  The nuclear reaction rates (p-p chains,
CNO tri-cycle, Ne--Na and Mg--Al chains and the most important
$\alpha$-capture reactions including the $\alpha$-n reactions) and the
corresponding $Q$-values are taken from the recommended rates in the
JINA reaclib data base \citep{Cyburt_etal10}. Electron screening factors
for all reactions are from \citet{Dewitt} and
\citet{Graboske}. Finally, electron neutrinos energy losses are
computed following \citet{Munakata1985}, \cite{Itoh83} and
\cite{Haft}.

The metal abundances, 0.0001$\leq$Z$\leq$0.04, follow the partition of
\cite{GS98} (GS98) with the recent revision by \cite{Caffau2011}. With
this revision, the current observed solar metallicity is $Z_{\odot}=
0.01524$.  The initial Helium content at varying metallicity is determined
by $Y=Y_{\rm p }+1.78\times~Z$, with $Y_{\rm p }=0.2485$ taken from
\cite{Komatsu_et_al_2011}.

\subsection{Convection, overshooting and mixing}
The mixing length parameter~\citep{BV1958}, derived from the solar
model, is $\alpha_{\rm MLT}=1.74$.  We adopt the Schwarzschild
criterion~\citep{Schwarzschild_criterion} to test the stability of radiative zones against convection.
In the presence of a gradient of chemical composition, an alternative
criterion is that of Ledoux~\citep{Ledoux_criterion}. This condition may happen during the evolution of massive stars
when the convective core grows in mass or when an intermediate radiative region
of varying  chemical composition becomes unstable to the convection.  In
this paper, we opt for the Schwarzschild criterion because, on one
side it has been shown that it is the more appropriate one to account
for the effects of thermal dissipation \citep{Kato1966} and, on the
other, the presence of a sizable overshooting region from the
convective core significantly reduces the differences between models
computed with the two alternative criteria \citep{Meynet2000}.

Overshooting from the convective core is estimated within the
framework of the mixing-length theory, allowing for the penetration of
convective elements into the stable regions \citep{Bressan_etal81}.
The adopted mean free path of convective elements {\em across} the
border of the unstable region, $l_{\rm c}$=$\Lambda_{\rm c}$\HP with
$\Lambda_{\rm c}=0.5$, is calibrated on the colour-magnitude diagram
of intermediate age clusters \citep{Girardi_etal09} as well as on
individual stars \citep{Kamath_etal10, Deheuvels_etal10, Torres_etal14}.
\begin{figure}
\centering
\includegraphics[scale=0.415]{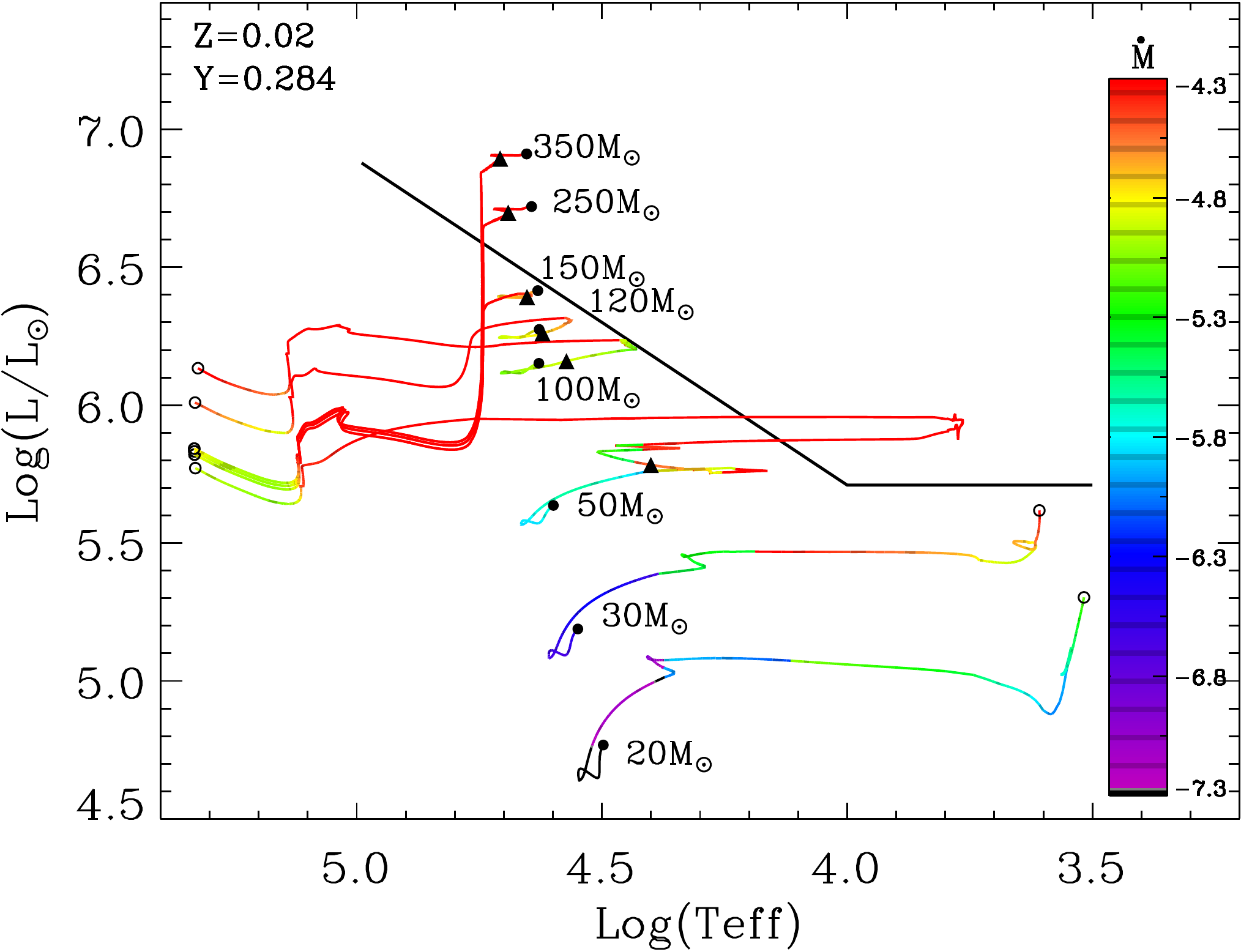}\\
\includegraphics[scale=0.415]{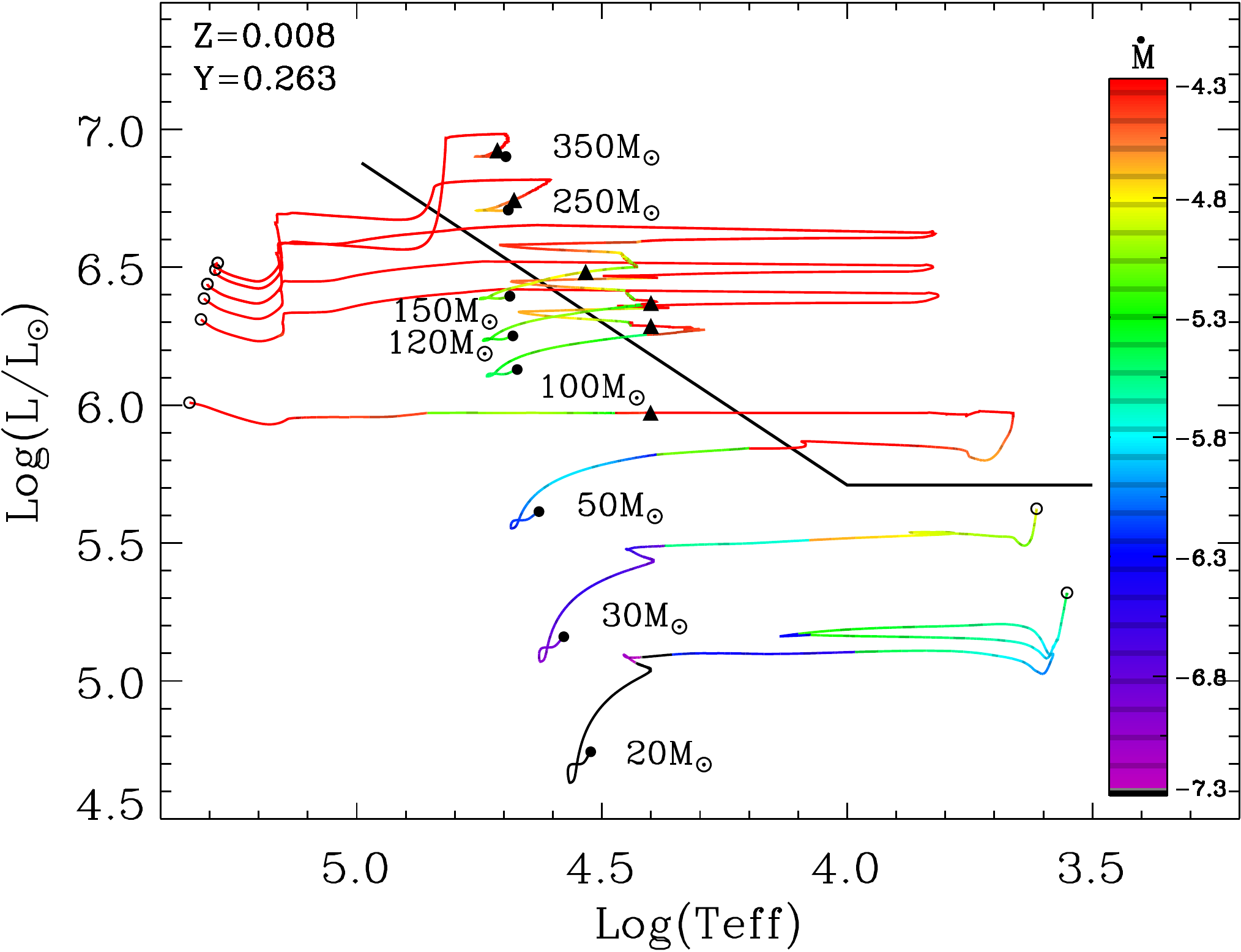}\\
\includegraphics[scale=0.415]{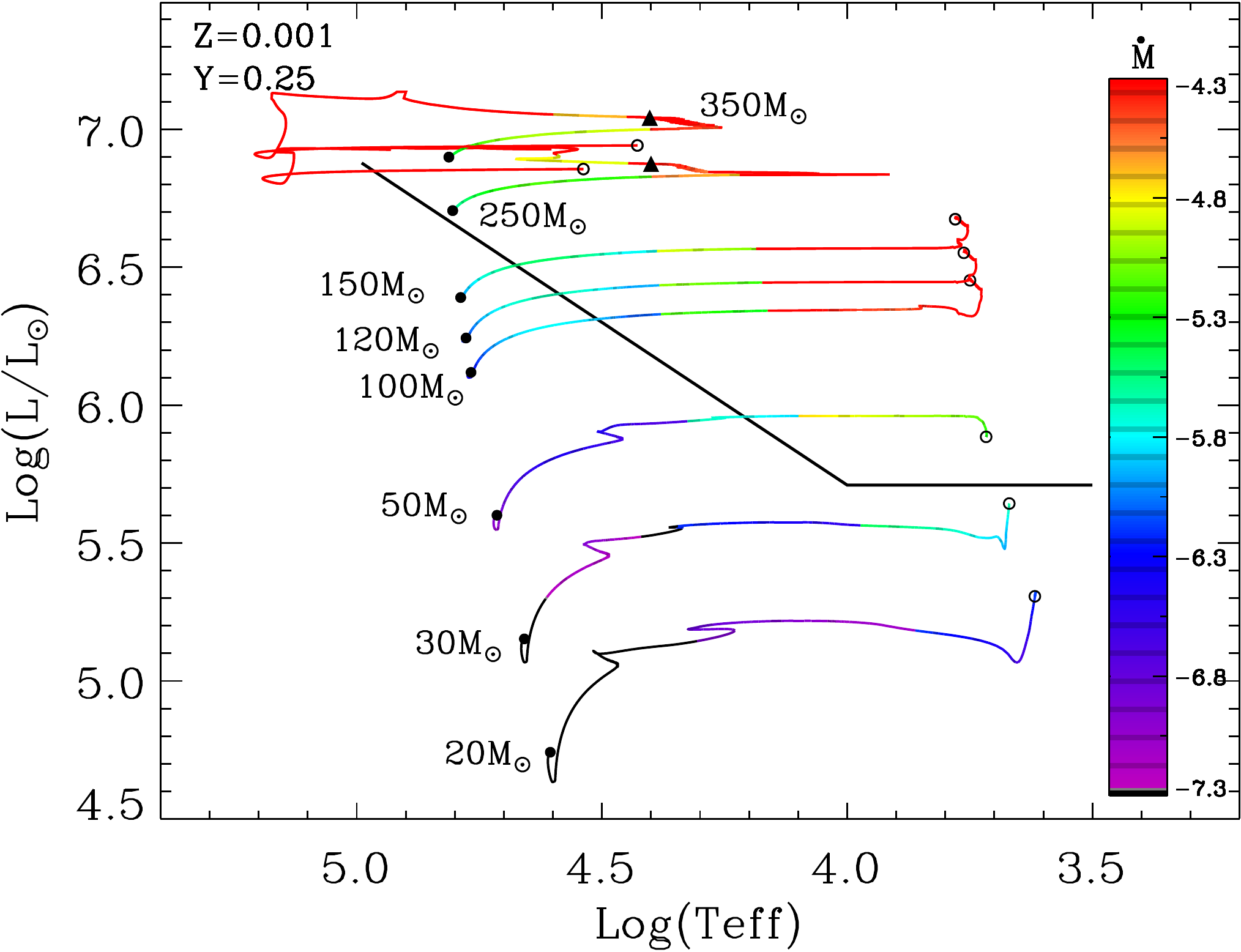}
\caption{Selected evolutionary tracks for massive stars with Z=0.02
  (upper panel), Z=0.008 (middle panel) and Z=0.001 (lower panel).
  The mass-loss rates are indicated with
  the colour bar. The thick black straight lines represent the
  Humphreys-Davidson limit~\citep{Humphreys_Davidson_1979} which
  delimits the forbidden region above which only very few
  stars are observed in the Hertzsprung-Russell (HR) diagram
  of the Galactic massive stars. The big solid and empty circles indicate the ZAMS and the end points of the tracks respectively.
  The triangles mark the beginning of WR phase.
\label{massive_tracks}}
\end{figure}
We also account for overshooting at the base of the convective envelope,
which is simply modelled by mixing the radiative region down to a
distance of $l_{\rm e}$=0.7\HP from the formal Schwarzschild border \citep{Alongi_etal91}.
We stress that the extent of the overshooting regions
and the corresponding mixing efficiencies are still a matter of debate.
Concerning core overshooting, a recent analysis of the period spacing of gravity modes
in low mass Helium burning stars, suggests a quite sizable overshooting region, $\Lambda_{\rm c}=1.0$
in the above formalism (Bossini et al. 2015, in preparation).
Concerning envelope overshooting, work in progress (\citet{Tang_etal2014} and Rosenfield et al. in preparation)
already indicates that using larger values of $l_{\rm e}$ (close to 2~\HP)
at the bottom of the convective envelope fits better
the extended blue loops seen in metal-poor dwarf galaxies.
Therefore, future releases of the PARSEC data base are likely to have these prescriptions revised.

Finally, rotational mixing has not yet been introduced in PARSEC.

\begin{figure}
\includegraphics[scale=0.415]{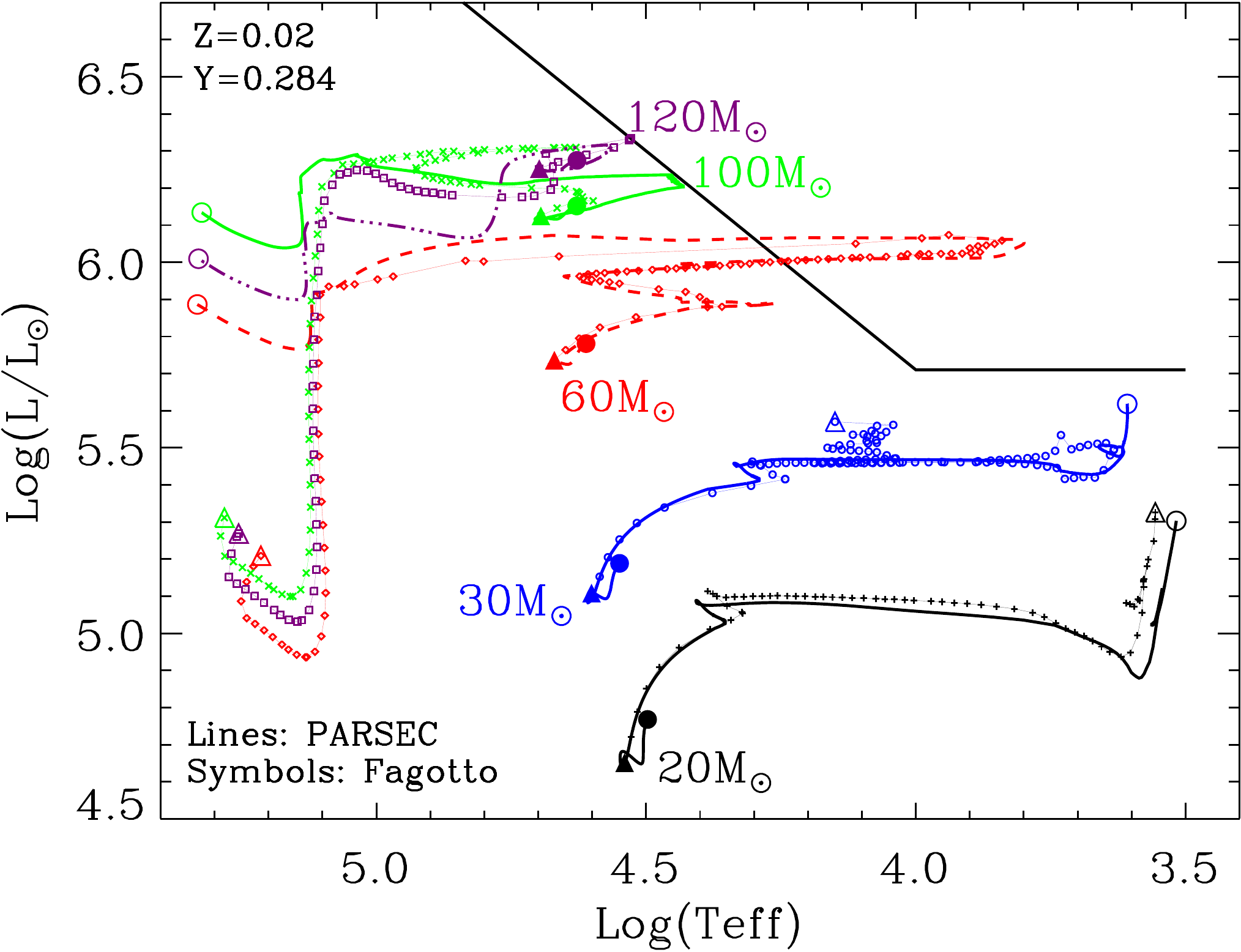}
\includegraphics[scale=0.415]{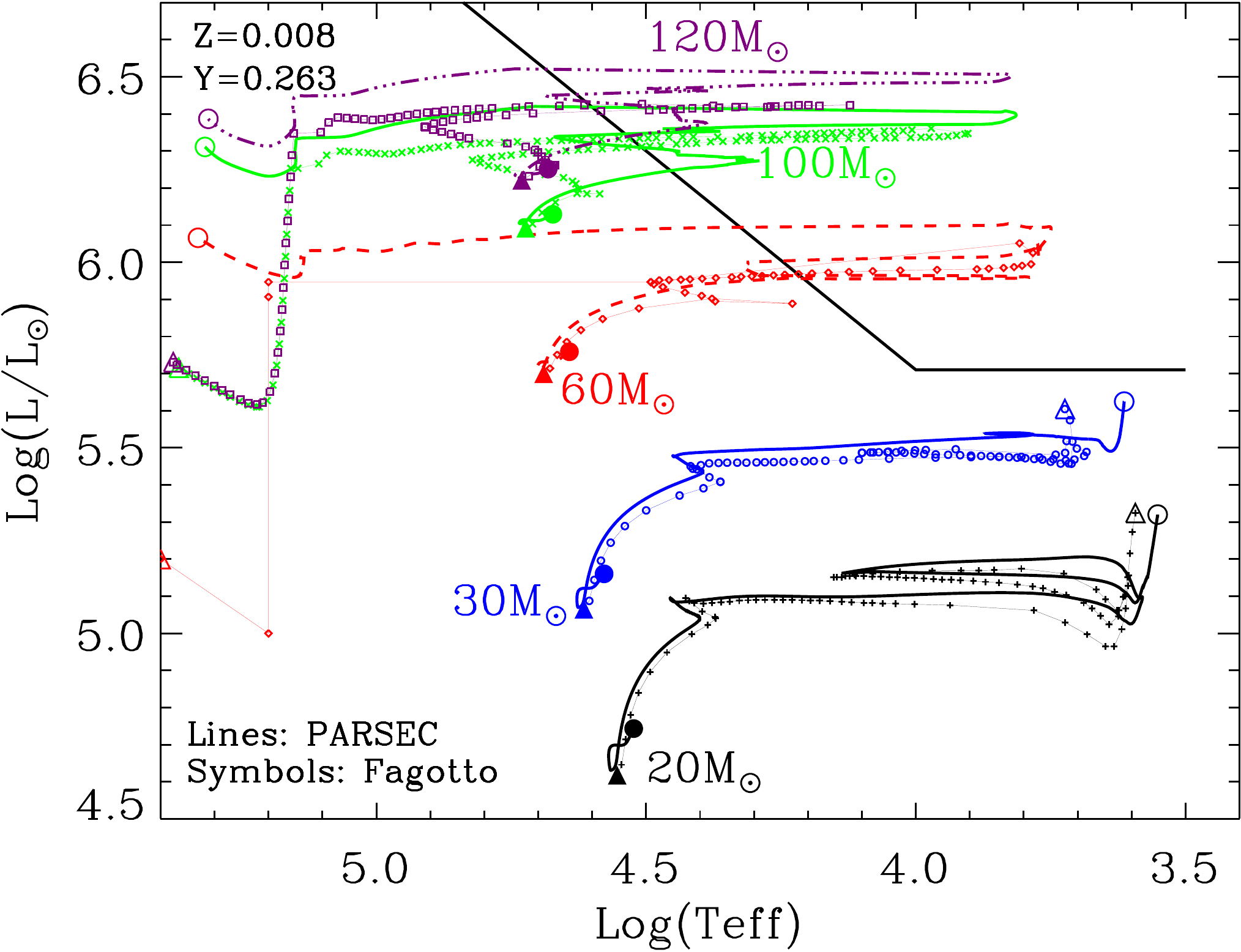}
\includegraphics[scale=0.415]{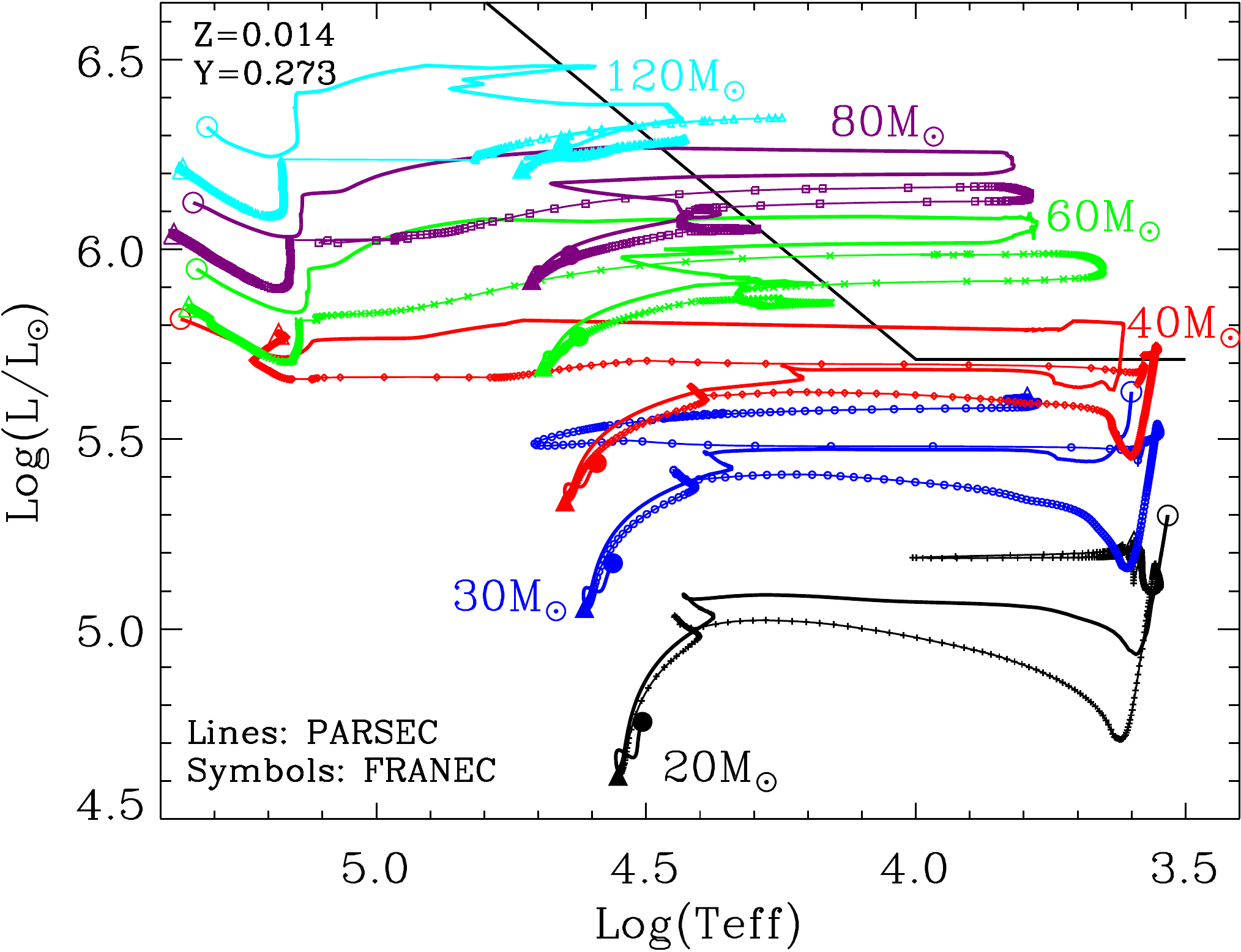}
\caption{Upper and middle panels: comparison with previous releases of
  Padova evolutionary tracks~\citep[symbols,][]{Fagotto_etal94,
    Bressan1993} with $Z=0.02$ (upper panel) and $Z=0.008$ (middle panel).  Models
  of different masses are indicated with different colours and line styles/symbols.
    Note that the Helium abundances are $Y=0.28$ for $Z=0.02$ and $Y=0.25$ for $Z=0.008$ in
  \citet{Fagotto_etal94, Bressan1993}.
  \hspace{0.3cm}
  Lower panel:
  comparison with the FRANEC solar abundance~($Z=0.01345$) models
  without rotation~\citep[symbols,][]{Chieffi_etal2013}. The meaning of big circles is the same as in figure \ref{massive_tracks}, while the big triangles are used for the alternative models. The Humphreys-Davidson limit is also drawn as in figure~\ref{massive_tracks}.}
\label{fig_comparison}
\end{figure}

\subsection{Mass-loss rates}
At solar metallicity, the mass-loss phenomenon is known to dominate
the evolution of stars for initial masses \Mini$\geq{30}M_\odot$.  We
account for this process following recent prescriptions
found in the literature for the different spectral types.  In the blue
supergiant phase, for $\teff \geq$ 12000\,K, we adopt the relations
provided by~\citet{Vink2000,Vink2001}.  This formulation (R$_{V01}$)
shows an almost linear overall dependence of the mass-loss rates on
the metallicity, $\dot{M}\propto(Z/Z_\odot)^{0.85}M_\odot/yr$.  In the
supergiant phases with  $\Teff <$ 12000\,K we use the mass-loss rates
provided by \citet{deJager1988}, R$_{dJ}$, assuming the same
dependence on the surface metallicity of R$_{V01}$.  For
WR stars we use the \cite{Nugis2000} formalism, R$_{NL}$. They also
provide a dependence on the stellar metallicity. The definition of the
WR phases is provided below (see section~\ref{hot_stars}).

An aspect which is relevant for the more massive stars concerns the
transition between the O-phase, the LBV/RSG phase and finally the WR
phase and, most importantly, the dependence of this transition upon
the metallicity of the stars.  For example, in the old Padova models,
e.g., \citet{Bressan1993}, the transition to the super-wind phase
corresponding to the LBV stars is artificially set at the stages when
the models cross the Humphreys-Davidson instability limit
\citep{Humphreys_Davidson_1979} in the HR diagram.
This is justified by the evidence that Galactic and Magellanic Clouds massive stars
near this limit show mass-loss rates that may reach $\dot{M}\simeq$10$^{-3}M_\odot/yr$.
However, while the Humphreys-Davidson limit is an observed property
of the HR diagram of massive stars in near solar environments,
it is used independently from the metallicity of the galaxy,
in spite of the fact that the mass-loss rates themselves do depend on the abundance
of heavy elements~\citep{Kudritzki2000, Puls2000, Mokiem2007, Smith2014}.
This approximation becomes critical at very low metallicities.

From the theoretical side, recent detailed studies of radiative wind models
(\citet{Grafener2008} and \citet{Vink2011}) show that the mass-loss
rates are strongly enhanced when the stars approach the electron
scattering Eddington limit
\begin{equation}
   \Gamma_e = \frac{L\kappa_{es}}{4\pi c G M}=1
\label{gamma}
\end{equation}
Since for the most massive stars at solar metallicity this may happen
near the Humphreys-Davidson limit,  the above formalisms could provide a modern description
of the transition from O-type through LBV/RSG-type to WR-types \citep{Vink2012}.
We thus include in  PARSEC the recent formulation of
mass-loss rates by \citet{Vink2011}.
The resulting HR diagram of a few selected evolutionary tracks for about solar
metallicity is shown in the upper panel of
figure~\ref{massive_tracks}. For purposes of clearness, we do not plot
the pre-main sequence phase.  The colours along the tracks represent the
strength of the mass-loss rates as indicated in the inset scale. In the figure, the
thick black lines mark the so-called Humphreys-Davidson
limit~\citep{Humphreys_Davidson_1979} which delimit the forbidden
region above which only very few stars are observed in the HR diagram
of the Galactic massive stars.  As shown in the figure, the main
sequence of the most massive stars extends up to the
Humphreys-Davidson limit. Eventually, the stars may encompass this
limit but the time spent in this region is very short because the mass-loss
rate becomes so high that the stars rapidly lose their envelopes
and turn into the hotter region of the HR diagram.  We stress that
this is not a result of an {\sl ad hoc} assumption for the mass-loss rate,
but a direct result of the application of the new adopted relations of
the mass-loss rates. Near the Humphreys-Davidson limit, $\Gamma_e$
rises close to 1 and, as described in \citet{Vink2011}, when $\Gamma_e$
is larger than 0.7, the mass-loss dependence on $\Gamma_e$ becomes
high and, correspondingly, the mass-loss rates are significantly
enhanced.  With this formulation, the boosting of the mass-loss rate
at the highest masses (\Mini$\geq150M_\odot$) is effective already from
the beginning of the main sequence and, because of the large mass-loss
rates, they evolve almost downward vertically in the HR diagram.
Interestingly, the luminosity of the tracks with the higher masses
(\Mini$\geq150M_\odot$) falls with time much more than those of the less
massive ones. This is caused by the {\sl over-luminosity} with respect
to the main sequence mass -- luminosity relation which, being larger
at larger masses, results in larger values of $\Gamma_e$.  Thus, the
evolved brightest massive stars are not necessarily those with the
largest initial masses.

The $\Gamma_e$ dependence of the mass-loss rates of O-type supergiant
stars has not yet been studied for different galactic environmental
conditions, apart from the analysis of the effects of CNO abundances
\citep{Muijres2012a}.  A more thorough analysis in a broad metallicity
range, $10^{-3}~Z_\odot\leq~Z\leq~2~Z_\odot$, has been performed by
\citet{Grafener2008}, but only for the case of WR stars.  In
particular, \citet{Grafener2008} show that the dependence of mass-loss
rates on the metallicity is also a strong function of
$\Gamma_e$. While at low values of $\Gamma_e$ the mass-loss rates obey
the relation of $\dot{M}\propto(Z/Z_\odot)^{0.85}M_\odot/yr$, at
increasing $\Gamma_e$ the metallicity dependence decreases, and it
disappears as $\Gamma_e$ approaches 1.  In the absence of a more
comprehensive analysis of the dependence of the mass-loss rates on the
metallicity and $\Gamma_e$, and since there is a continuity between
the models provided by \citet{Vink2011} and those of WNL stars
provided by~\citet{Grafener2008} (see discussion in \citet{Vink2011}),
we assume in PARSEC that the scaling with the metallicity obeys the
following relation
\begin{equation}
\dot{M}\propto~(Z/0.02)^\alpha
\label{mdotgamma}
\end{equation}
with the coefficient $\alpha$ determined from a rough fit to the
published relationships by \citet{Grafener2008}:
\begin{equation}
   \alpha = 2.45-2.4*\Gamma_e \qquad (2/3\leq\Gamma_e<{1})
\label{mdotalpha}
\end{equation}
and with the supplementary condition $0<\alpha\leq{0.85}$.
In summary, our algorithm for the mass-loss is the following.  Besides
the already specified mass-loss rate formulations (R$_{V01}$, R$_{dJ}$
and R$_{NL}$) we compute also R$_{\Gamma_e}$ from the tables provided
by \citet{Vink2011}, but we scale the latter value with the
metallicity, using equations (\ref{mdotgamma}) and (\ref{mdotalpha}).
During the BSGs and  LBVs phases the basic mass loss rate adopted is
R$_{V01}$. However, since, because of the effects of $\Gamma_e$,
this can be encompassed by R$_{\Gamma_e}$ and in order to secure a smooth transition,
we adopt the maximum between R$_{V01}$ and R$_{\Gamma_e}$.
Toward the red supergiant phase we use R$_{dJ}$, but again, since this is an empirical rate parameterized
in such a way that it should  hold on a broad region of the HR diagram and it likely underestimates
the mass loss rates of luminous yellow super-giants \citep{Salasnich1999},
we compare it with R$_{\Gamma_e}$ and take the maximum value.
In the WR phases we consider the \citet{Nugis2000} formulation.

The HR diagram of a few selected evolutionary tracks for Z=0.008 is
shown in the middle panel of figure~\ref{massive_tracks}.  We note here
a significant decrease of the mass-loss rates, at a given mass.  In
particular while at Z=0.02 the models of \Mini$\sim$120M$_\odot$ and
\Mini$\sim$150M$_\odot$, rapidly turn their main sequence evolution to
higher effective temperatures, at Z=0.008 the mass-loss rates are not
high enough to prevent the tracks from entering into the forbidden
region.  Nevertheless, the tracks burn Hydrogen around the
Humphreys-Davidson limit until, near central Helium ignition $\Gamma_e$
becomes large and after performing a rapid excursion within the
forbidden region, they turn into the blue part as WR stars. At even
lower metallicities, the mass-loss rates decrease, unless the star is
near the Eddington limit with $\Gamma_e\sim$1, and the location of the
predicted Humphreys-Davidson limit shifts to higher luminosities.  For
example, at Z=0.001 in the lower panel of figure~\ref{massive_tracks},
the upper main-sequence widens significantly and the
more massive stars evolve into the ``forbidden'' region even during
the H-burning phase, because of their very large convective
cores. They may also ignite and burn central Helium as ``red''
supergiant stars.

\subsection{Comparison with previous releases}
\label{sec_comparison}

The current set of massive stars supersedes the old one adopted in
several popular studies since 20 years \citep[][hereafter BF+]{Bressan1993, Fagotto_etal94}.
With respect to BF+, the new models
have a significantly finer mass spacing and extend up to a higher
upper mass limit.  The computed masses range from 14 to 20\,\Msun in steps of
2\,\Msun, then up to 100\,\Msun in steps of 5\,\Msun and finally 120, 150,
200, 250, 300, 350\,\Msun.  This allows for a quite better
interpolation in ages and masses and a sampling of different initial
mass function up to larger initial masses.  Simple stellar populations
can now be sampled with a higher accuracy than before and with more
suitable mass and time-steps.  The new tracks include also the
pre-main sequence phase that begins when the central temperature of
the protostar becomes larger than log(T$_c/K)=5.3$. No mass accretion
is accounted for during the pre-main sequence phase.

To summarize the differences brought by the adoption of the updated
physics input we compare the new tracks with those of the previous
Padova release, in the upper and middle panels of
figure~\ref{fig_comparison}, for $Z=0.02$ and $Z=0.008$ respectively.
Besides the presence of the pre-main sequence in the new tracks, which for purposes
of clarity is not shown, a few general trends can be seen in the
figure.  At Z=0.02, the Zero Age Main Sequence (ZAMS) is similar.
Note that the old tracks do not include the pre-main sequence phase. The
main sequence termination is only slightly hotter in the new tracks,
likely because of differences in the underlying opacities. The red
supergiant phase is slightly cooler in the new tracks, but the
differences are barely significant.
We remind that in BF+ the density
inversion (arising in the inefficient region of the convective
envelope) was inhibited in the computation of massive star tracks.
In a later revision of the tracks \citep{Girardi_etal00} it was inhibited
at all masses. As discussed in \citet{Alongi1993},
a density inversion may develop in the external inefficient convection zones
because of the requirement of the hydrostatic equilibrium in a region with
a large super-adiabatic real temperature gradient (assuming typical values of the
mixing length parameter). This situation which should lead to a Rayleigh-Taylor
instability or even a significant increase in the mass-loss rate, may give rise
to numerical instabilities which preclude the computation of the track.
To inhibit the density inversion one may use a mixing length parameter proportional
to the density scale height ($H_\rho$) which, by  rendering convection more efficient in the region where
the density has a relative maximum, prevents a large super-adiabatic real temperature gradient.
Alternatively one may impose that the real temperature gradient is limited by
$\nabla_{T}\leq\nabla_{T_{\rm max}}=\frac{1-\chi_\mu\nabla_\mu}{\chi_T}$,
which is the choice made since BF+. This also simulates a more efficient convection and prevents the development of numerical instabilities.
However since a larger convective efficiency results in a hotter red giant track, and since
BF+ red giant tracks of intermediate and low mass stars compared well with the corresponding
observed colours, we follow their method and inhibit density inversion only for the most massive stars.
Thus the most massive red super-giants  have effective temperatures that are slightly hotter than
those obtained by allowing density inversion to occur.
A detailed analysis of this effect is clearly needed, but likely it must also take into account effects
of dusty circumstellar envelopes which are known to affect the colours of RSG stars.

The mass-loss rates adopted
here are less efficient already at \Mini=30\,\!\Msun.  At this mass, the star
ignites Carbon as a red giant while in BF+ the star is already moving
toward the hot region of the HR diagram.  This effect is more striking in
the more massive stars, for those reaching the WR stages.  Comparing
the two models of 60\,\Msun we see that while the H-burning and the
central Helium ignition phases are pretty similar, the final WR phase is
very different.  The mass-loss adopted for the WR stages in BF+ is
significantly larger than that adopted here and the star ends with a
luminosity which is about one order of magnitude lower.

\begin{figure}
\centering
\includegraphics[scale=0.48]{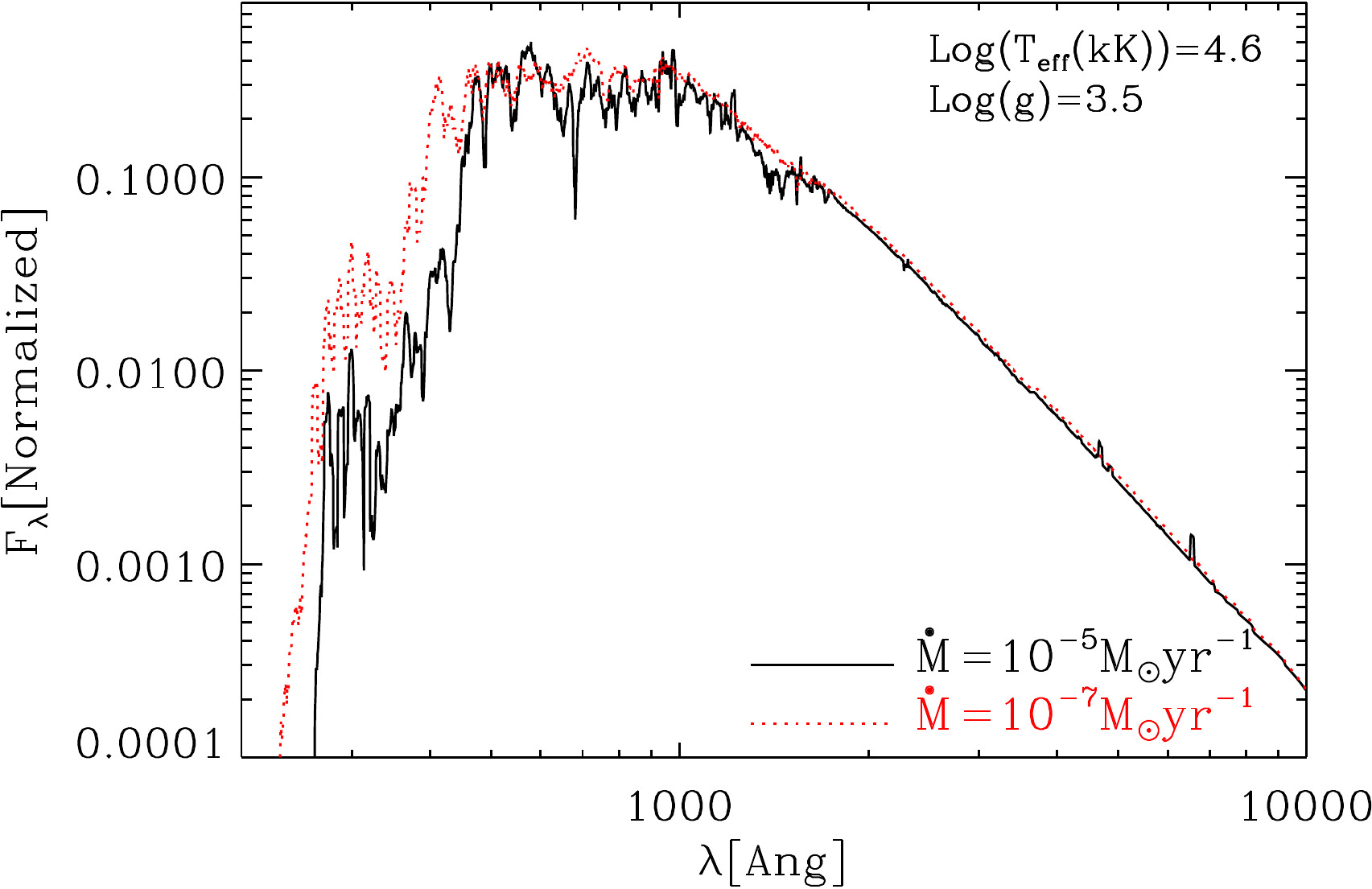}
\includegraphics[scale=0.48]{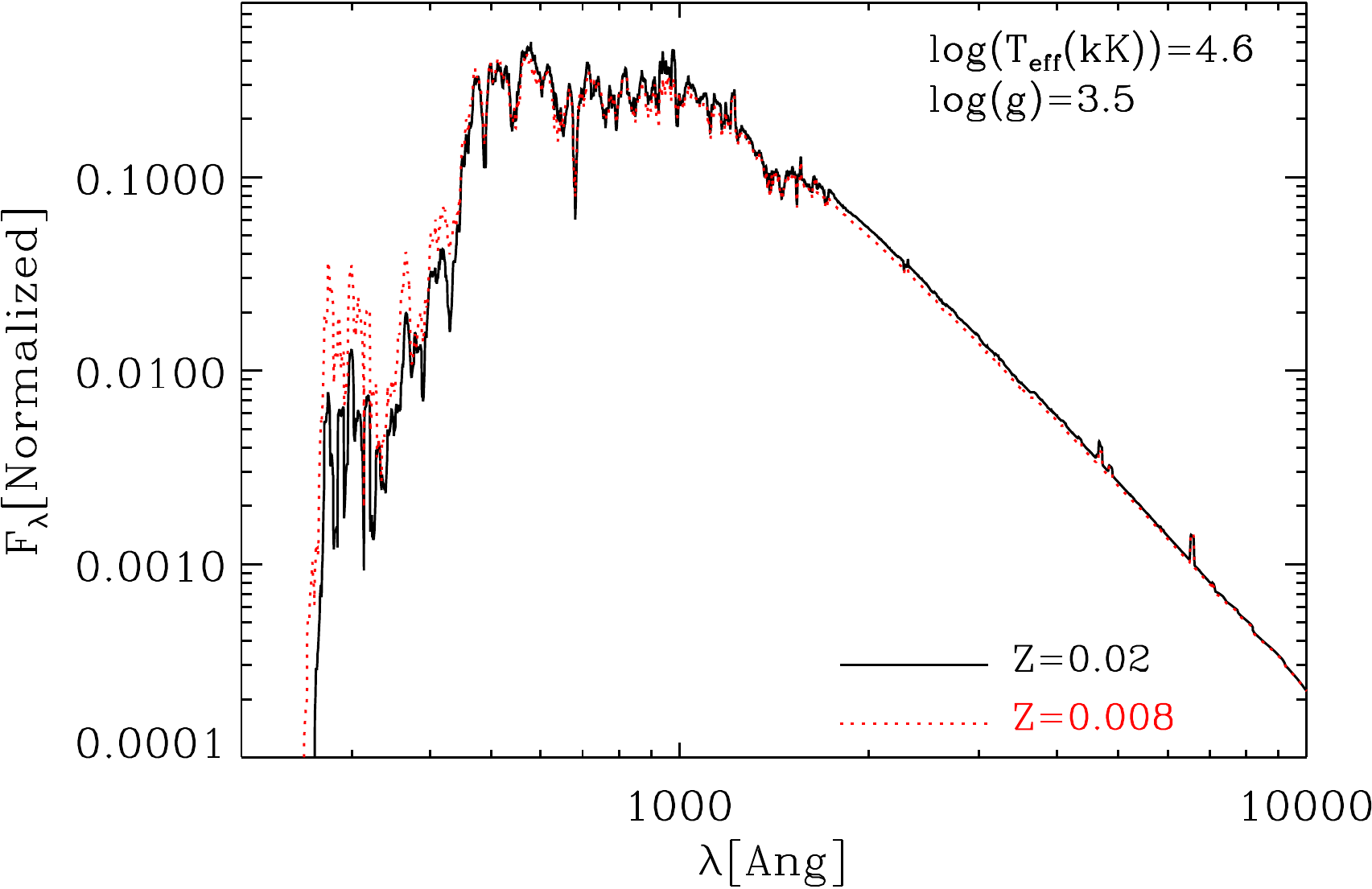}
\caption{Upper panel: comparison of {\sl WM-basic} spectra (smoothed)
  with different mass-loss rates. The effective temperature, gravity,
  radius and metallicity ($Z=0.02$) are the same, but the black solid line is for
  $\dot{M}=10^{-5}M_\odot\ yr^{-1}$ and the red dotted line is for
  $\dot{M}=10^{-7}M_\odot\ yr^{-1}$. The ionizing photon (with
  wavelength less than $912~{\rm nm}$) number ratio ($N_{\rm
    Ionizing}(\dot{M}=10^{-5}M_\odot\ yr^{-1})/N_{\rm
    Ionizing}(\dot{M}=10^{-7}M_\odot\ yr^{-1})$ supposing the same
  luminosity) is 0.197.\hspace{.5cm} Lower panel: comparison of
  {\sl WM-basic} spectra (smoothed) with different metallicities. The
  effective temperature, gravity, radius and mass-loss rate
  ($\dot{M}=10^{-5}M_\odot\ yr^{-1}$) are the same, but the black solid line is
  for $Z=0.02$ and the red dotted line is for $Z=0.008$. The ionizing photon
  number ratio ($N_{\rm Ionizing}(Z=0.02)/N_{\rm Ionizing}(Z=0.008)$
  supposing the same luminosity) is 0.994 (almost identical).
  All the spectra are normalized over the wavelength range from 0.84 to 1.44\,$\micron$.}
\label{WMbasic-comparison}
\end{figure}
At higher masses, \Mini=100 and 120\,\Msun, there are significant differences
already in the core H-burning. The mass-loss rates are initially
higher in the BF+ models, but when the effects of $\Gamma_e$ become
important, the mass-loss rates become comparable and the tracks evolve
along the similar path. There remain a large difference in the
mass-loss rate adopted in the advanced WR stages.

We stress again that in BF+ models the mass-loss is arbitrarily
enhanced when the stars approach or encompass the Humphreys-Davidson
limit, while in the current models the mass-loss enhancement is a
result of the photospheric conditions.  In this respect, it is very
interesting to note the similarity between the tracks of \Mini=60\,\Msun
that can be taken as representative of the fact that the mass-loss
enhancement is no more imposed as before, but it is instead a natural
consequence of the {\sl photospheric} conditions.

The comparisons with the models of Z=0.008, typical of the Large
Magellanic Cloud, confirm the previous picture. The ZAMS is pretty
similar, but now the effects of a lower metallicity become visible,
since the normalization metallicity for the mass-loss rate has been
assumed to be Z=0.02.
In the old models, the metallicity dependence
was slightly lower, $\dot{M}\propto~(Z/0.02)^{0.5}$, and the mass-loss
rates at Z=0.008 were about 40\% larger than in the current
models, keeping other parameters fixed.  Thus, the new models
evolve at slightly higher luminosity until the effects of $\Gamma_e$
become important. Again there is a striking similarity in the turnover
of the effective temperature in the models that overcome the
Humphreys-Davidson limit. As before, in the final WR stages the stars
evolve at a significantly higher luminosity than that of the old
models.  We note that, while in the advanced WR stages the luminosities of
the new models are significantly higher, the corresponding lifetime is
more or less unchanged ($\sim 0.1\,{\rm Myr}$ for $100\,\Msolar$
models).  This means that the new models should contribute far more to
the hard ionizing photons than the old ones. This will be the subject
of a forthcoming paper dedicated to the integrated properties of young
star clusters.

\begin{figure*}
\centering
\includegraphics[scale=0.4]{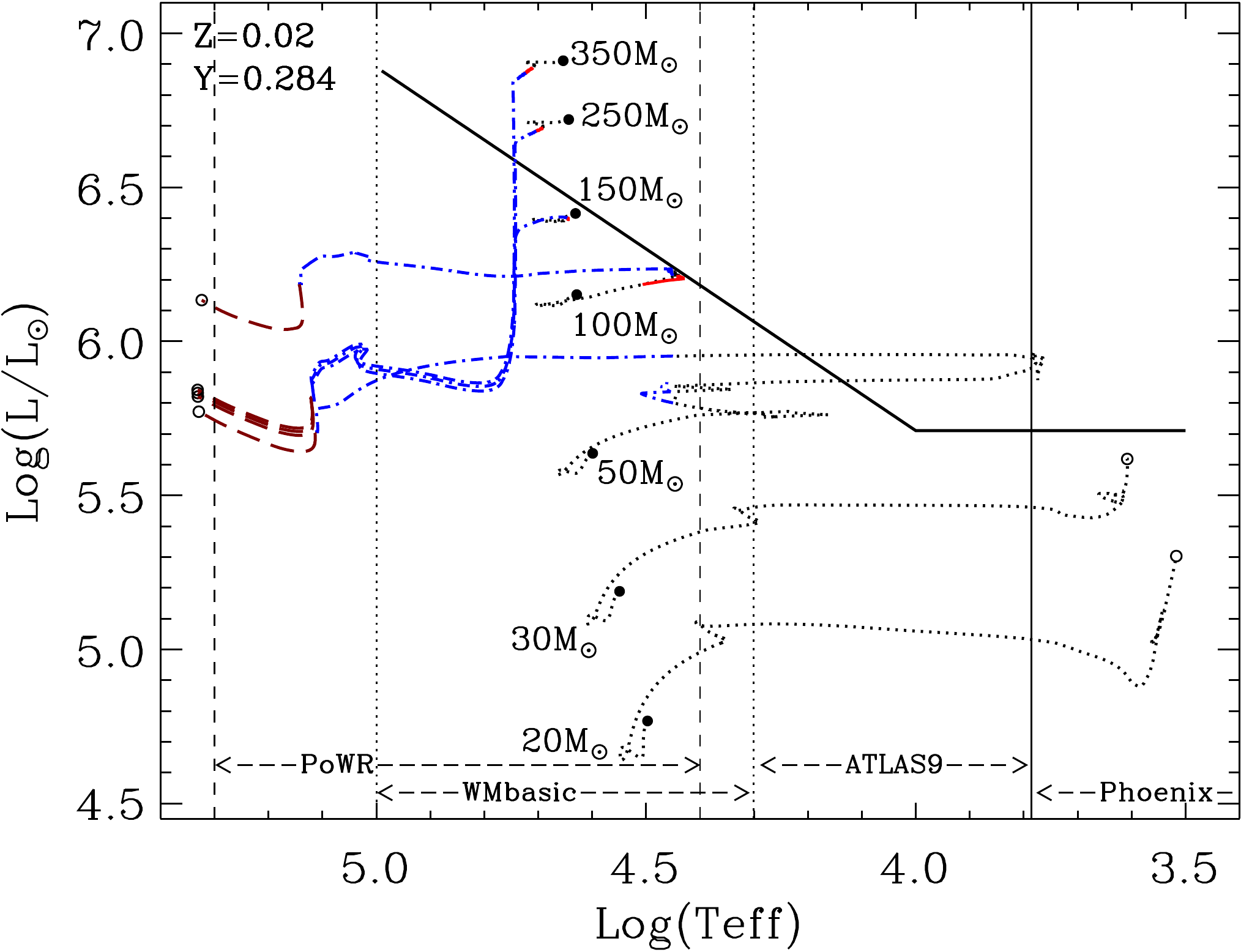}~
\includegraphics[scale=0.4]{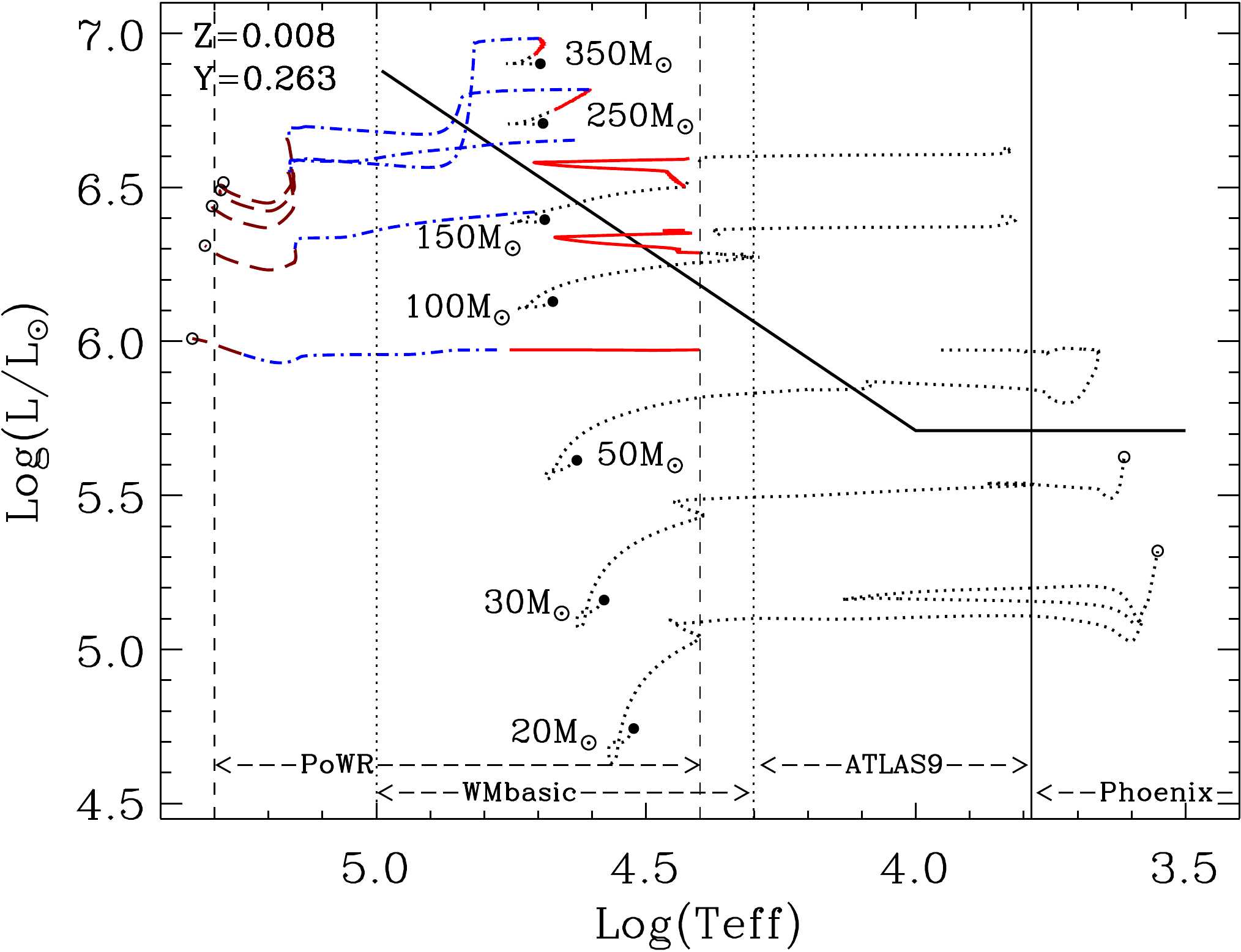}
\caption{Evolutionary tracks for massive stars with Z=0.02 (left panel) and
  Z=0.008 (right panel). Different colours represent different evolutionary
  stages: black dotted lines for stages precedent of WR phases, red solid lines for models using PoWR WNL-H50
  ($Z=0.02$) or WNL-H40 ($Z=0.008$), blue dash-dotted lines for WN models, and brown dashed lines for WC
  models. The meaning of big circles is the same as in figure \ref{massive_tracks}.
  The overplotted vertical lines delimit the coverages of
  different atmosphere models as indicated in the
  plots.  The Humphreys-Davidson limit is also drawn as in figure \ref{massive_tracks}. \label{massive_tracks_conlib}}
\end{figure*}
For the comparison with models by other groups, we select
the evolutionary tracks without rotation
computed with the FRANEC code \citep{Chieffi_etal2013}.
The HR diagram  is shown in the lower panel of
figure~\ref{fig_comparison}.  The FRANEC models (dashed lines) are for
a composition of $Z=0.01345$ and $Y=0.265$ while PARSEC models (solid
lines) are for a slightly different chemical composition,
$Z=0.014$ and $Y=0.273$.
They adopt a mixing-length parameter $\alpha \equiv\Lambda/H_p=2.3$ and a core
overshooting region of $0.2H_p$
\citep{Chieffi_etal2013,Martins&Palacios2013}.  One of the major
difference is the presence of big blue loops in the FRANEC $20\,, 30$
and $40\,\Msolar$ models, while the big blue loops are absent in our
models.
In this respect,
we note that the main sequence termination of our models is
significantly cooler and more luminous than that of FRANEC
models. This is particularly evident in the model of \Mini=$40\,\Msolar$
and it is indicative of a larger central mixing during the main
sequence phase of our models, which is known to disfavour the
occurrence of blue loops.
The larger extension of the loops in the FRANEC models could also be due
to their use of the Ledoux criterion for the definition of the borders of the
intermediate unstable regions. It is well known that the Ledoux criterion inhibits the development of
the intermediate convective regions within the Hydrogen profile, at the
end of the H-burning phase, favouring a deeper penetration of the
convective envelope when the stars move into the RSG region.
This is an important issue deserving further investigation.
Preliminary models computed
adopting either the Schwarzschild  or the Ledoux criterion in the
Hydrogen chemical composition profile, confirm that the main
cause of the inhibition of blue loops is the extent of core overshooting (or another kind of extended mixing)
during main Hydrogen burning phase. With a slightly reduced core overshooting,
also PARSEC models with the Schwarzschild criterion show extended blue loops (Tang et al. 2015 in preparation).

We also see that the behavior of the tracks within the
Humphreys-Davidson region is very similar.  In both cases, the models
abandon this forbidden region after losing a significant fraction of
their mass.  In both cases, the mass-loss must be enhanced, but the
invoked mechanism are different: \citet{Chieffi_etal2013} invoke the dust
mass-loss enhancement \citep{vanLoon2005}, while we invoke the
$\Gamma_e$ mass-loss enhancement.

\section{Stellar atmosphere models}

We convert from effective temperature, gravity and luminosity to
magnitudes and colours using theoretical atmosphere models.  Since
one of our aims is to provide also isochrones in the observational plane, we
need a full set of atmosphere models encompassing a wide range of
parameters, i.e. masses, evolutionary stages and metallicities.  In
the following, we describe the different atmosphere models adopted. We begin with
ATLAS models \citep{Kurucz1993, Castelli2004}, which are the canonical
models suitable for intermediate and low mass stars.   We
complement this library with our new calculations of wind models
for hot massive stars with the {\sl WM-basic} code~\citep{WMbasic},
and new models of WR stars from the Potsdam
group~\red{\citep{PoWR1, PoWR2, PoWR3,PoWR4,PoWR5,PoWR6}}.
For the coolest stars we adopt the Phoenix models \citep{Allard1997}, already
extensively discussed in our previous paper \citep{Chen2014}.

\subsection{Intermediate mass and low mass stars}
The core of our library consists of the plane parallel ATLAS9
models \citep{Kurucz1993}.  These LTE (Local thermal equilibrium)
models are well suited to describe the atmospheres of intermediate
and low mass stars of spectral type between A and K.  The most recent
ATLAS9 models are those computed by \cite{Castelli2004}.
ATLAS9 models are based on the solar abundances by \cite{GS98} and
make use of an improved set of molecular lines including TiO, $\rm
H_2O$, HI-Hi and $\rm HI-H^+$. The model grids are computed for
$\teff$ from 3500 K to 50000 K, $\logg$ from 0.0 dex to 5.0 dex and
[M/H]=+0.5, +0.2, 0.0, -0.5, -1.0, -1.5, -2.0, -2.5, -3.5, -4 and
-5.5.  We limit the use of ATLAS9 models to the temperature range of
19000\,K$>\teff>$6000\,K. At higher temperature in general we need to
consider models with mass-loss while at lower temperature the Phoenix
models are more appropriate.

\subsection{Hot massive stars}
\label{hot_stars}
\noindent {\bf O, B stars.}\qquad For temperatures typical of O and B
stars (60000\,K$>\teff>$19000\,K) we have generated a new library of
models using the public code {\sl WM-basic} \citep{WMbasic}. This
allows us to take into account both the effects of extended
winds and those of non-LTE, since  both effects may significantly affect
the emergent spectra of hot stars.
We have generated new sets of stellar spectral libraries
covering as much as possible the space of parameters of our new
evolutionary tracks of massive stars, i.e.  effective temperature,
gravity, metallicity and mass-loss rate.  The temperature grid ranges
from $\logteff = 4.3$ to $5.0$ in steps of $\Delta \logteff=0.025$\,
dex.  At each temperature, we adopt a step in gravity of $\Delta
\logg=0.5$\,dex, with upper and the lower boundaries that depend on
the $\teff$.  Indeed, the highest gravity is determined by the fact
that the line driven radiation force cannot initiate the stellar wind,
while the lowest value corresponds to stability problems when the
models approach the Eddington limit.  At each grid point we consider
three different values of the mass-loss rates, $\dot{M}=10^{-7}$,
$10^{-6}$, and $10^{-5}\, M_{\odot}/yr$, encompassing typical values for O, B
stars.

\begin{figure*}
\centering
\includegraphics[angle=0,scale=0.49]{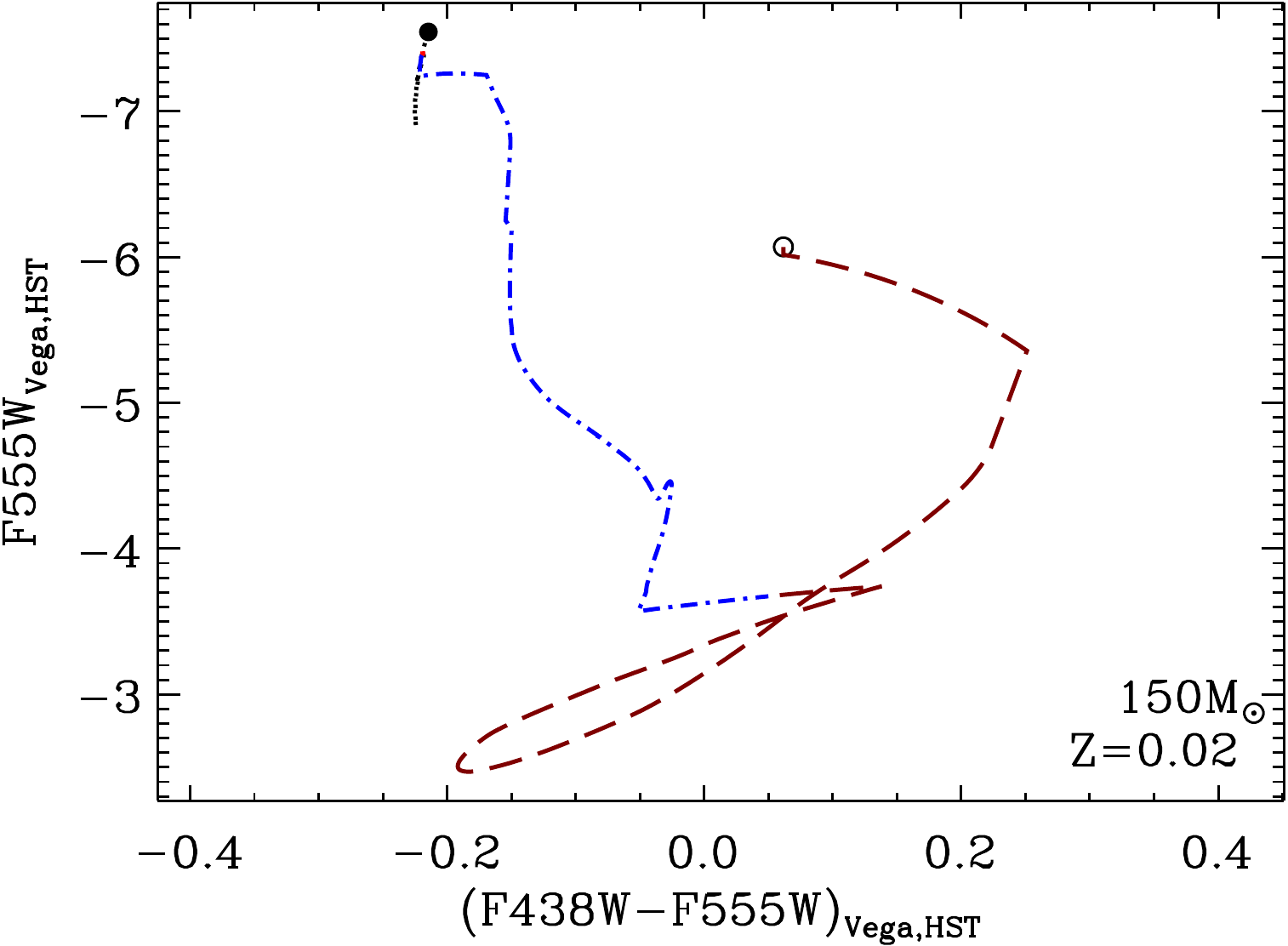}~
\includegraphics[angle=0,scale=0.49]{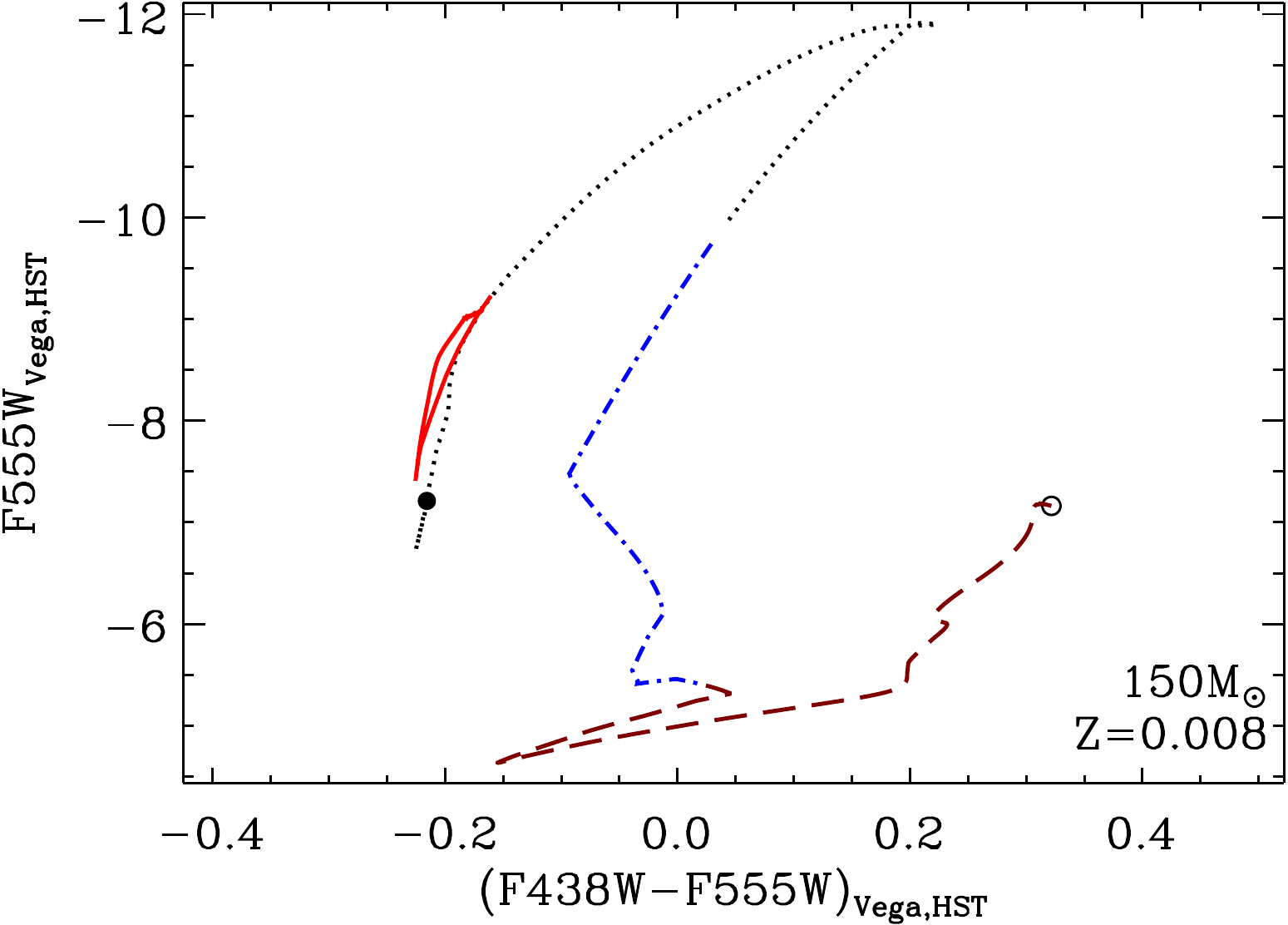}
\includegraphics[angle=0,scale=0.49]{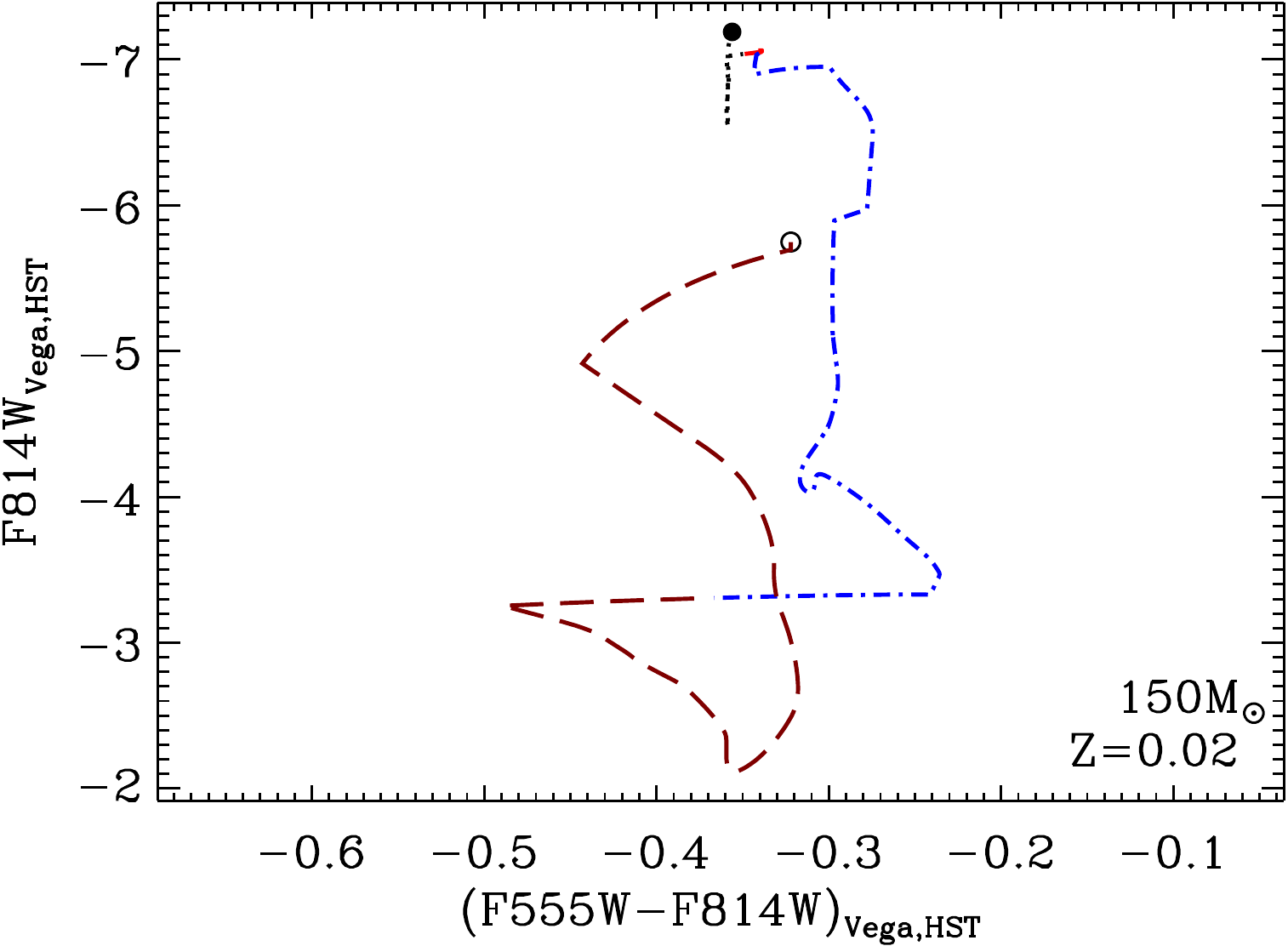}~
\includegraphics[angle=0,scale=0.49]{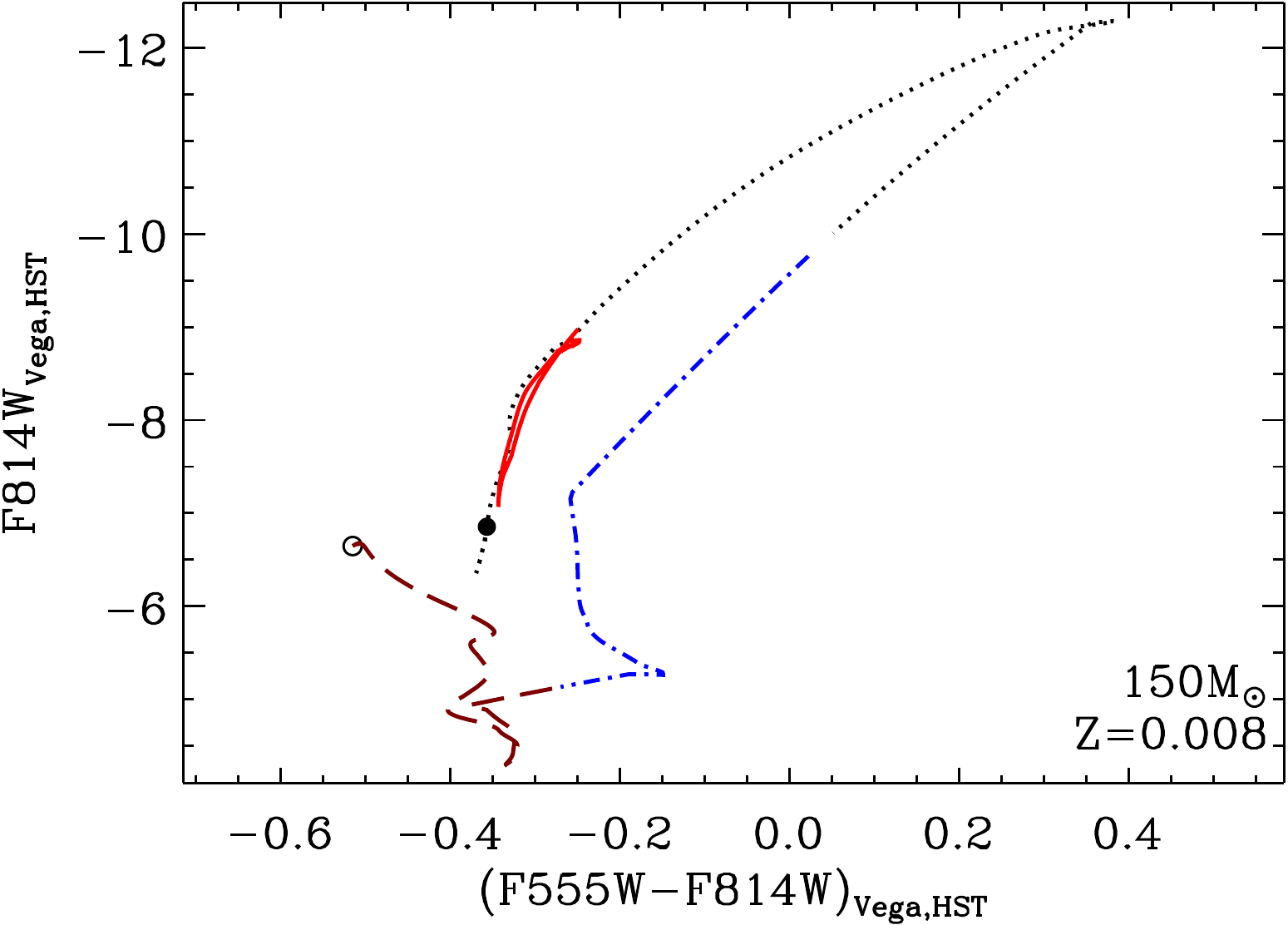}
\includegraphics[angle=0,scale=0.49]{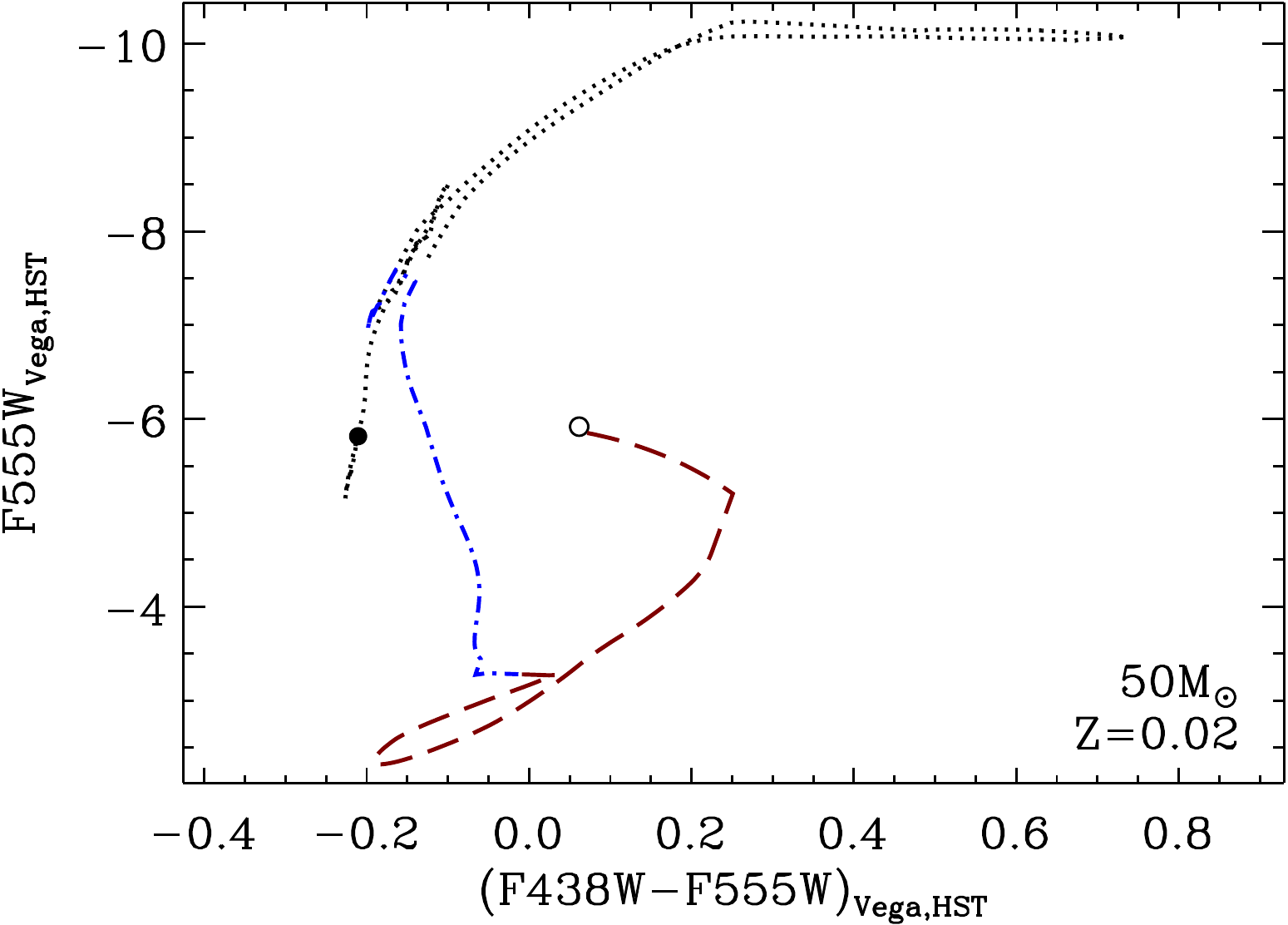}~
\includegraphics[angle=0,scale=0.49]{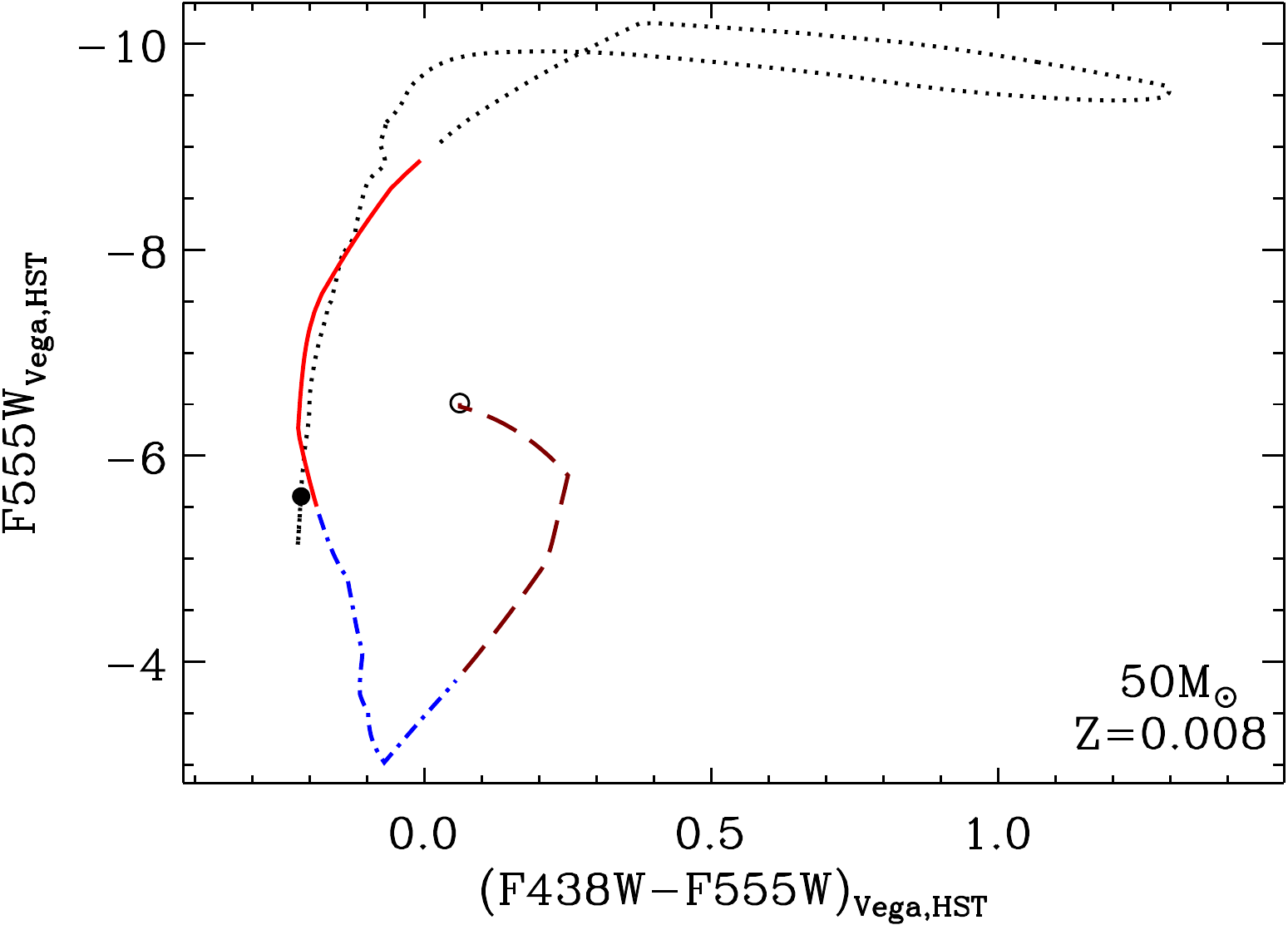}
\includegraphics[angle=0,scale=0.49]{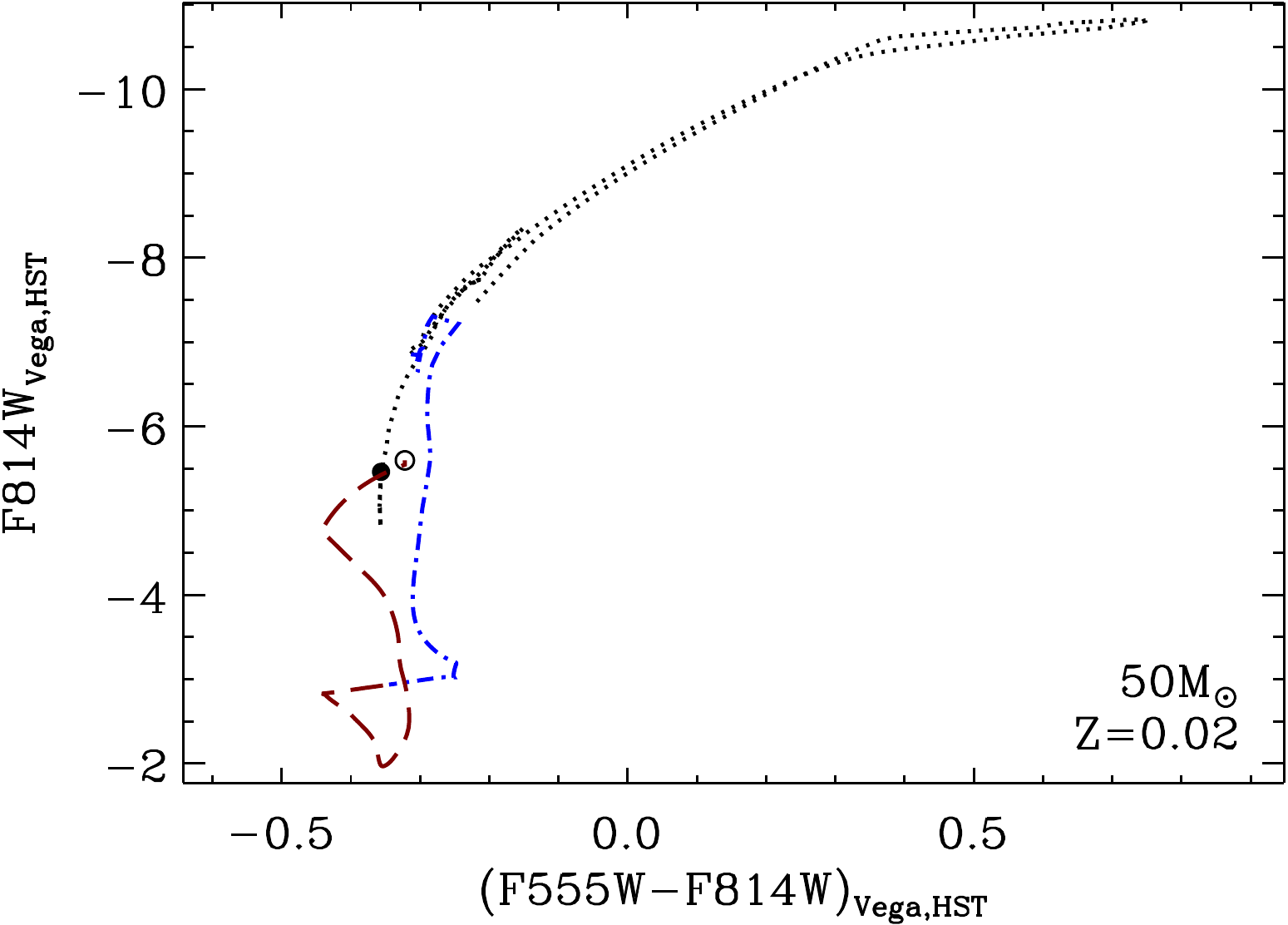}~
\includegraphics[angle=0,scale=0.49]{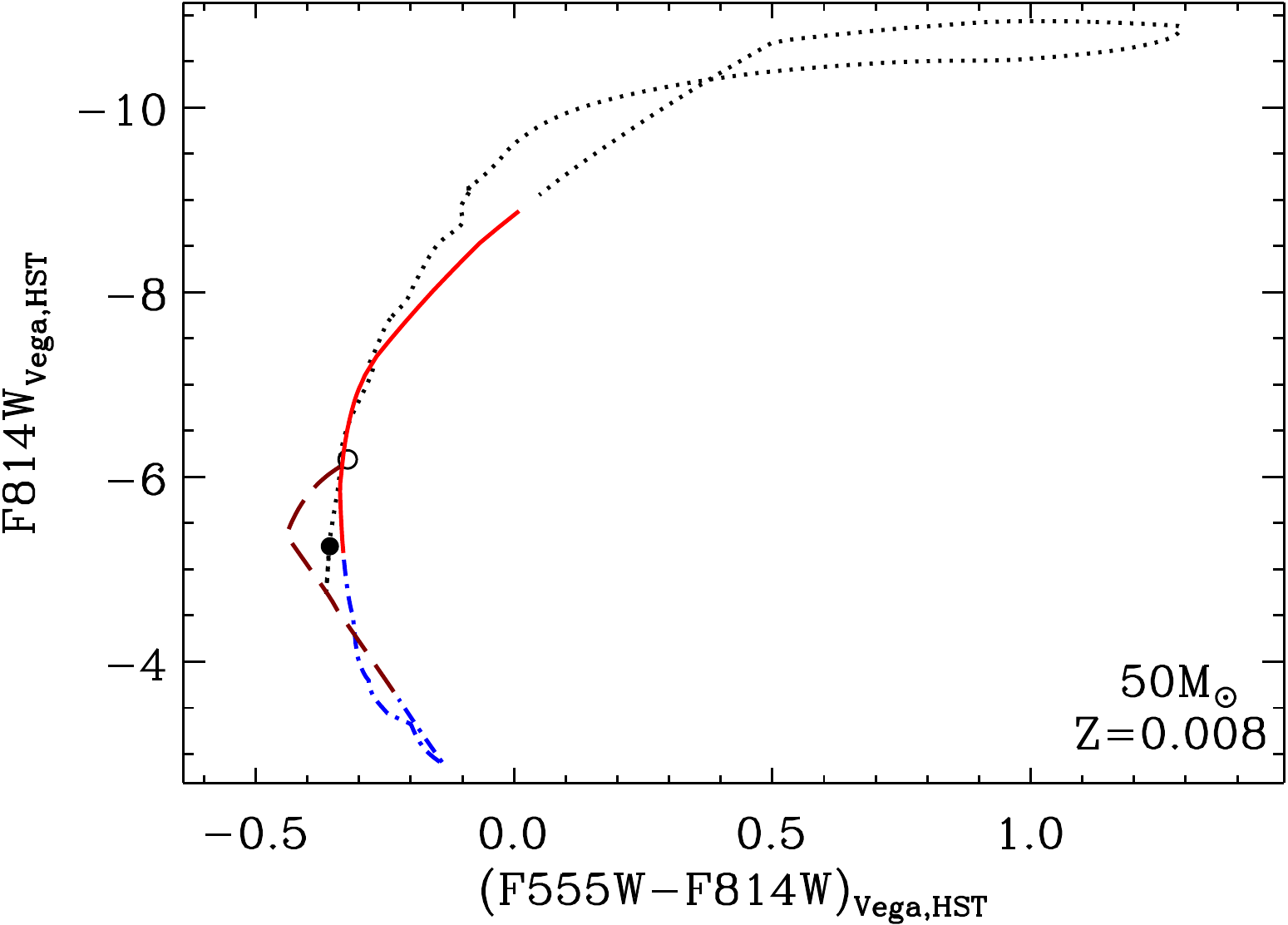}
\caption{Colour-magnitude diagrams in HST/WFC3 broad bands for tracks of \Mini=$150M_\odot$ (upper four panels)
  and \Mini=$50M_\odot$ (lower four panels) with
  Z=0.02 (left panels) and Z=0.008 (right panels) respectively. The meaning of colours
  and line styles along the tracks is the same as in
  figure~\ref{massive_tracks_conlib}. The meaning of big circles is the same as in figure \ref{massive_tracks}.}
\label{cmd_track}
\end{figure*}
We note that, besides $\teff$, $\logg$ and stellar radius (radius defined at
a Rosseland optical depth of $2/3$, $R_{*,2/3}$), there are additional
input parameters in {\sl WM-basic}, $\kappa$, $\alpha$ and $\delta$,
referring most specifically to the structure of the wind as shown by
\citet{CAK}.  Thus, for any point in the (effective temperature,
gravity) grid we suitably change the other stellar parameters (radius,
luminosity and mass) as well as the wind parameters $\alpha$ and
$\kappa$ in order to obtain three consistent atmosphere models with
the required values of mass-loss rate, $\dot{M}=10^{-7}, 10^{-6}, 10^{-5} M_{\odot}/yr$.
These models are used to derive the broad band colours and magnitudes
for any given effective temperature, gravity and
metallicity, easily interpolating for the different values of
the mass-loss rate actually used in the tracks.
Our new library consists of about 300 models for each metallicity. For
example for $z=0.02$, we computed 105 models for $\dot{M}=10^{-7}M_\odot\ yr^{-1}$, 98
for $\dot{M}=10^{-6}M_\odot\ yr^{-1}$ and 86 for $\dot{M}=10^{-5}M_\odot\ yr^{-1}$.  The metal partition
adopted in the new models is the same as used in \cite{PARSEC}, to say
\cite{Caffau2011}.  As an example, we show in
figure~\ref{WMbasic-comparison}, the effects of changing the mass-loss
rate at constant metallicity (upper panel) and those of changing the
metallicity at constant mass-loss rate (lower panel), for models with
$\logteff=4.6$ and $\logg=3.5$.  In the upper panel, the black solid line
refers to the model with $\dot{M}=10^{-5}M_\odot\ yr^{-1}$, while the
red dotted one is for $\dot{M}=10^{-7}M_\odot\ yr^{-1}$.  We can see that the
${\rm He\, II}$ continuum is more absorbed in the model with higher
mass-loss rate.  We notice also that, as expected, some spectral features which appears
in emission in the higher mass-loss rate model, turn into absorption
in the model with a lower mass-loss rate.

In the lower panel, the black solid line refers to the same model as in the
upper panel, while, the red dotted one is now for a model at the same grid
point but with $Z=0.008$.  This comparison shows that the effects of changing the
metallicity on these hot spectra are less pronounced than those
arising from the variation of the mass-loss rates.

\smallskip
\smallskip
\noindent {\bf Wolf-Rayet stars.}\qquad WR stars typically
have wind densities one order of magnitude larger than those of
massive O-type stars.  Spectroscopically they are dominated by the
presence of strong broad emission lines of Helium, Nitrogen, Carbon
and Oxygen.  They are subdivided into different sub-types, one with
strong lines of Helium and Nitrogen (WN stars), another one with strong lines of
Helium and Carbon (WC stars) and a third one with strong Oxygen lines (WO
stars).

To reproduce their spectra we make use of the most recent library of
WR models computed by the Potsdam group (PoWR)~\red{\citep{PoWR1,PoWR2,PoWR3,PoWR4,PoWR5,PoWR6}}.
They provide models with metallicities corresponding to those
of WR stars in the Galaxy, the LMC and the Small Magellanic Cloud (SMC).  The model grids are
parameterized with ${\rm T_{eff,\tau=20}}$ (the effective temperature at
radius where the Rosseland optical depth is 20), and the transformed
radius $R_t$ (because models with the same set of such parameters show
the same emergent spectrum as discussed in \cite{Schmutz1992}).  The transformed
radius $R_t$ is given by \red{\citep{Schmutz1989}}:
\begin{equation}
R_t=R^*_{\tau=20}\Biggl(\frac{v_\infty}{2500\,km\,s^{-1}}\frac{10^{-4}M_\odot
  yr^{-1}}{\dot{M}}\Biggr)^{\red{2/3}}.
\end{equation}
Since $R^*_{\tau=20}$ is nearly equivalent to the hydrostatic radius
of the evolutionary tracks, this quantity combines stellar radius and
mass-loss rate of the tracks with the terminal velocity.  For the same
reason ${\rm T_{eff,\tau=20}}$ corresponds to the effective
temperature of our hydrostatic models $\teff^*$.
As shown in
\citet{PoWR3}, at large $R_t$ (or thin winds), ${\rm T_{eff,\tau=20}}
\backsimeq {\rm T_{eff,\tau=2/3}}$. (In the case of our {\sl WM-basic}
models, $R_t$ are always $\gtrsim 1.5$, so there is no need to use
$R_t$ to match atmosphere models and theoretical tracks).  As done in
\cite{Schmutz1992}, we match the WR atmospheres to the
evolutionary tracks by interpolating $\teff^*$ and $R_t$.
\begin{figure}
\centering
\includegraphics[angle=180,scale=0.36]{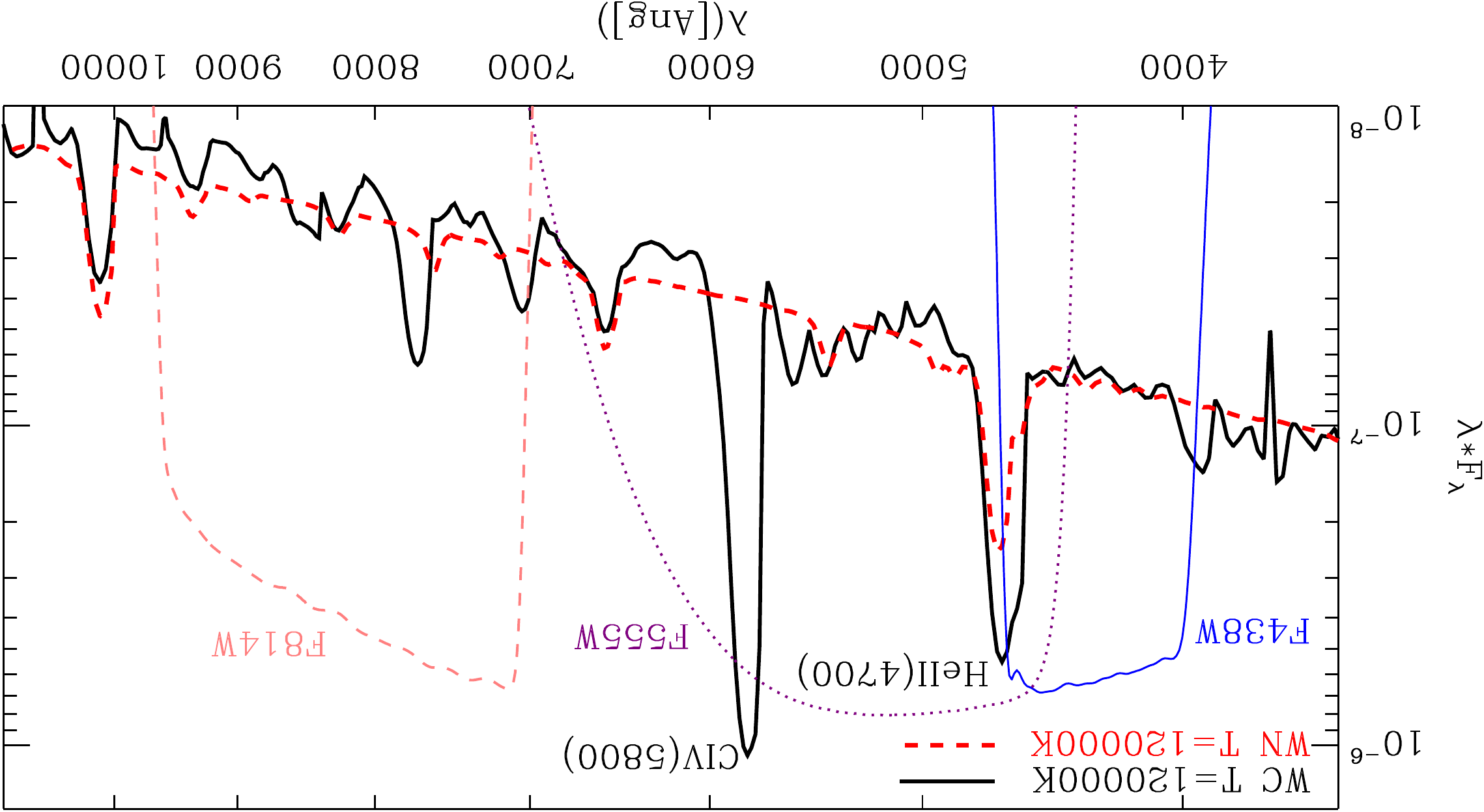}
\caption{Comparison of CMFGEN~\citep{CMFGEN} WC and WN
  models with the same effective temperature $\teff =120000\,K$ and
  metallicity Z=0.02.  Also shown are the transmission curves of three
  HST/WFC3 filters (F438W, F555W and F814W).
  Notice that the two strong emission lines HeII(4700) and CIV(5800) can
  greatly influence the magnitude/colours in related passbands. \label{compare_WC_WN}}
\end{figure}

While the observational classification is quite well
defined~\citep{WRreview, Crowther98}, it is more difficult to assign
the WR subgroups along the evolutionary tracks.  We use the following
convention to identify them.  WR stars are classified as WNL (late)
when the surface Hydrogen mass fraction $X_H$ is below a given threshold, X$_{\rm WNL}$.  When
$\rm X_H=0$, they are classified as WNE (early) if $N(^{12}{\rm C}) <
N(^{14}{\rm N})$ and as WC if $N(^{12}{\rm C}) \geq N(^{14}{\rm
  N})$. We finally assign the WO subtype when the condition
$N(^{12}{\rm C})+ N(^{16}{\rm O}) > N(^4{\rm He})$ is fulfilled
\citep{Smith1991}.  We then match our type assignments with the models
provided by PoWR, which adopt a fixed composition for any given
subtype.  For example, for the Galactic metallicity, in the PoWR
library X$_{\rm WNL}$=0.5 and the different subtypes have the following
surface compositions:
\begin{itemize}
  \item{\rm WNL-H50} for {\small $X_{\rm H}\!=\!X_{\rm WNL}$,
  $X_{\rm He}\!=\!0.48$, $X_{\rm C}\!=\!{\rm 1E\!-\!4}$, $X_{\rm N}\!=\!0.015$,
    $X_{\rm O}\!=\!0$ and $X_{\rm Fe}\!=\!0.0014$};
    \item{\rm WNL} for {\small
  $X_{\rm H}\!=\!0.2$, $X_{\rm He}\!=\!0.78$, $X_{\rm C}\!=\!{\rm 1E\!-\!4}$,
      $X_{\rm N}\!=\!0.015$, $X_{\rm O}\!=\!0$ and $X_{\rm Fe}\!=\!0.0014$};
      \item{\rm WNE}
for {\small $X_{\rm H}\!=\!0$, $X_{\rm He}\!=\!0.98$, $X_{\rm C}\!=\!{\rm
    1E\!-\!4}$, $X_{\rm N}\!=\!0.015$, $X_{\rm O}\!=\!0$ and $X_{\rm
    Fe}\!=\!0.0014$};
\item{\rm WC} for {\small $X_{\rm H}\!=\!0$, $X_{\rm
    He}\!=\!0.55$, $X_{\rm C}\!=\!0.4$, $X_{\rm N}\!=\!0$, $X_{\rm O}\!=\!0.05$ and
  $X_{\rm Fe}\!=\!0.0016$}.
\end{itemize}
For the LMC, the surface
Hydrogen threshold is X$_{\rm WNL}$=0.4, while for the SMC they provide models
with both X$_{\rm WNL}$=0.6 (WNL-H60 models) and X$_{\rm WNL}$=0.4 (WNL-H40
models).  Note that PoWR does not yet provide spectral models for WO
stars, which, however, are quite rare objects especially at lower
metallicities.

\begin{figure*}
\centering \includegraphics[scale=0.51]{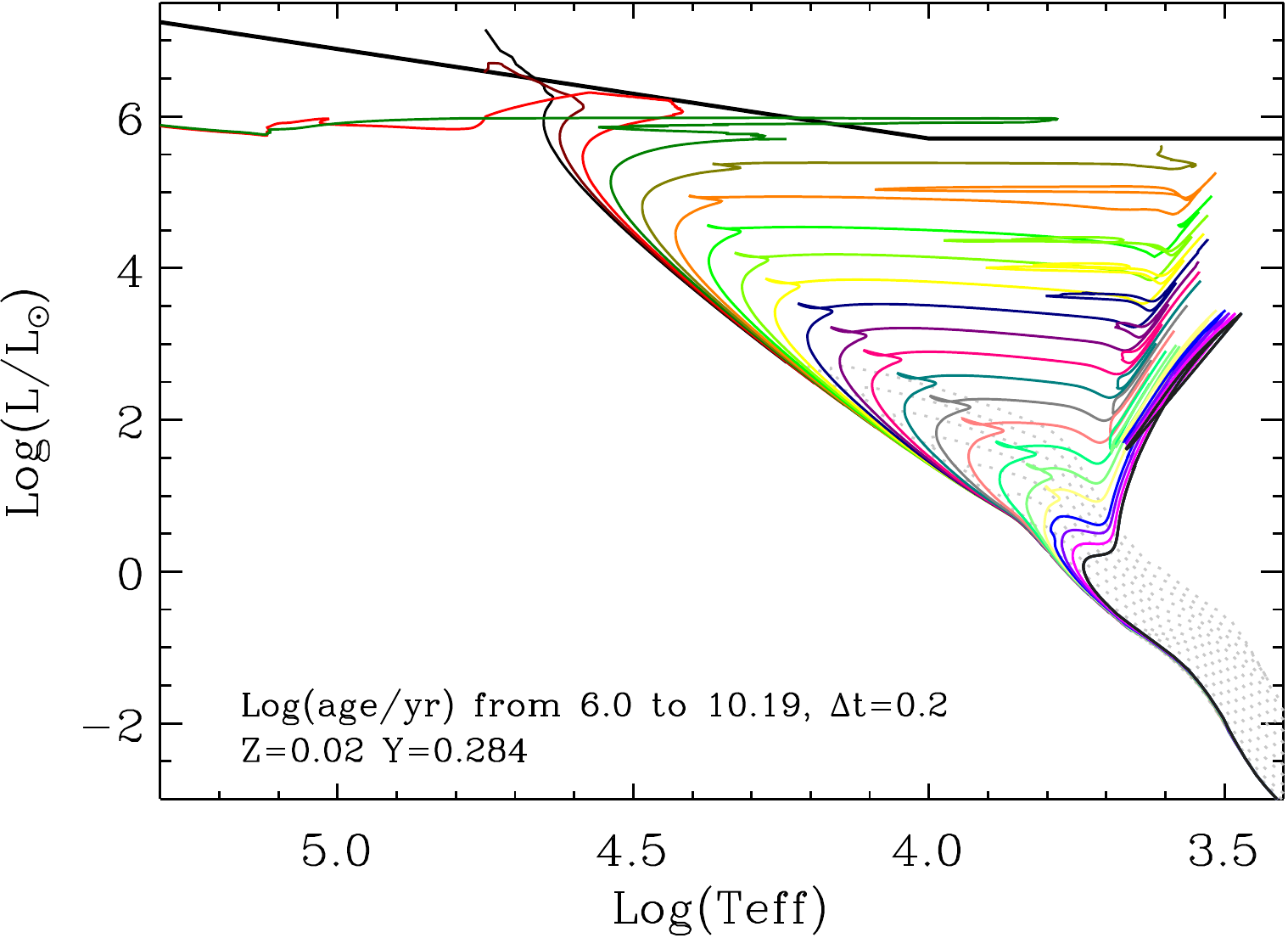}~
\includegraphics[scale=0.51]{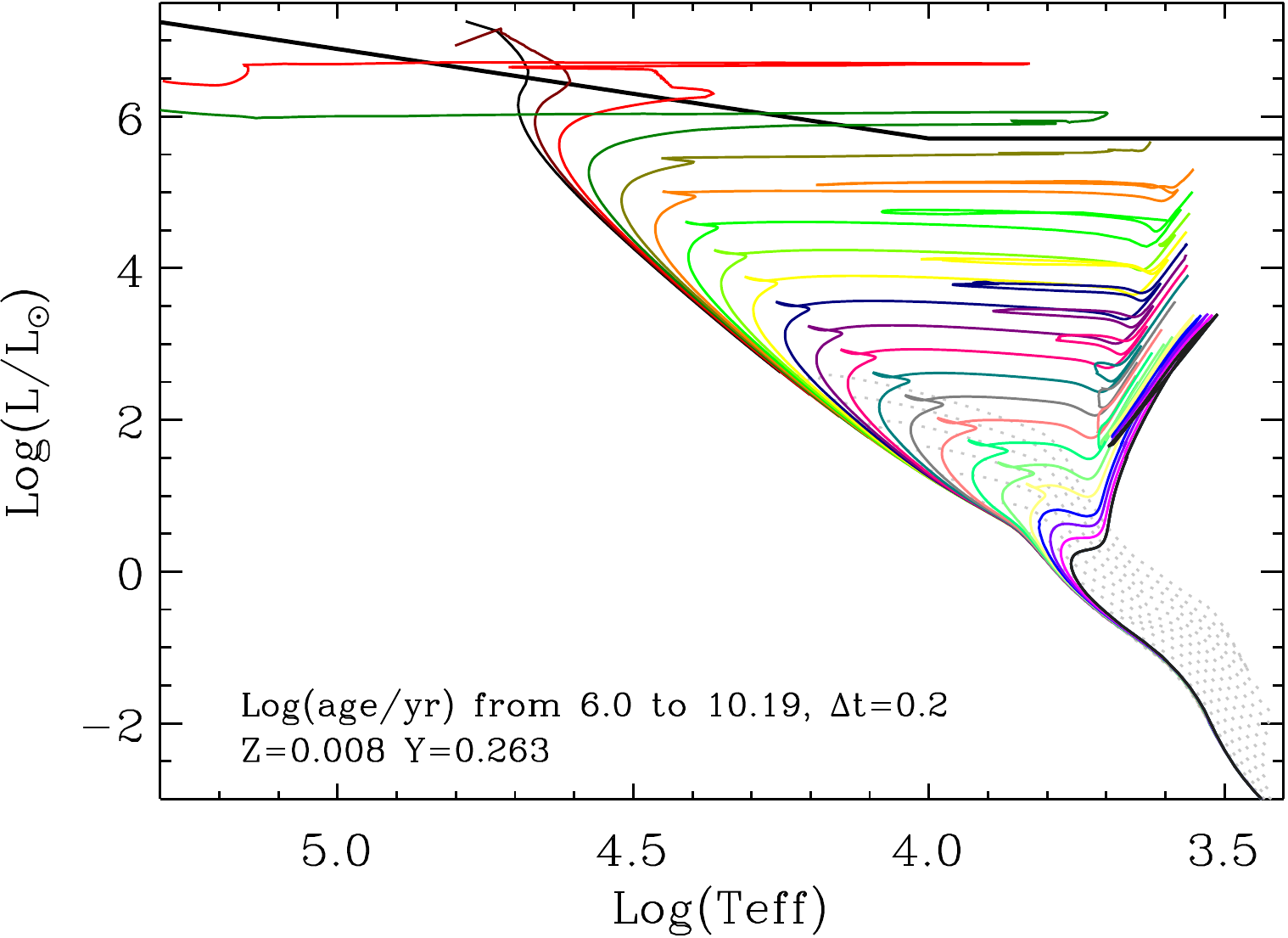}
\caption{Isochrones of different ages, as indicated by the labels, are
  shown for Z=0.02 (left panel) and Z=0.008 (right panel).  Note that at young
  ages the intermediate and low mass stars are still on the pre-main
  sequence (grey dotted lines).  The Humphreys-Davidson limit is also drawn as in figure \ref{massive_tracks}. \label{isochroneZ}}
\end{figure*}

\begin{figure*}
\centering
\includegraphics[scale=0.44,angle=0]{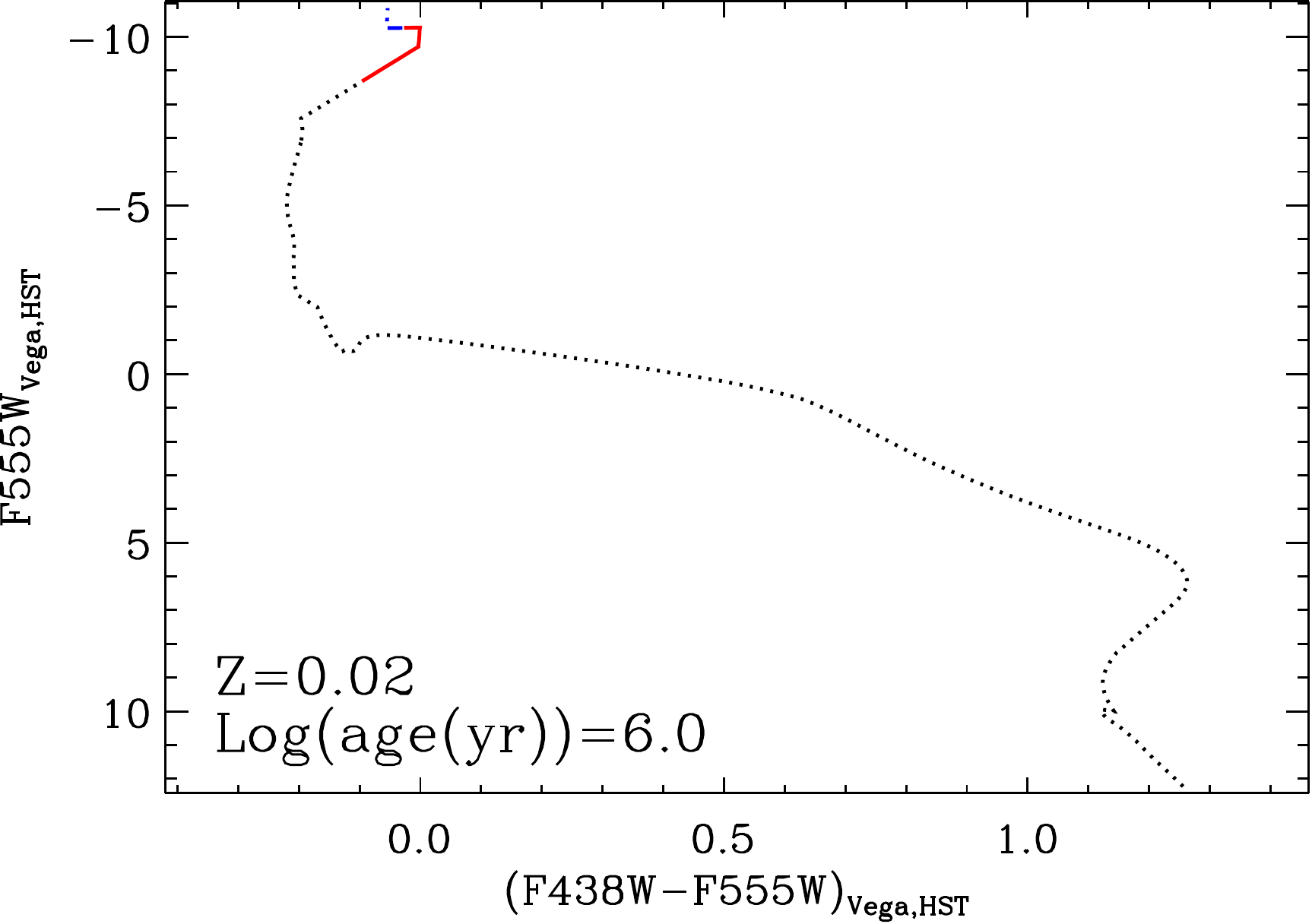}~
\includegraphics[scale=0.44,angle=0]{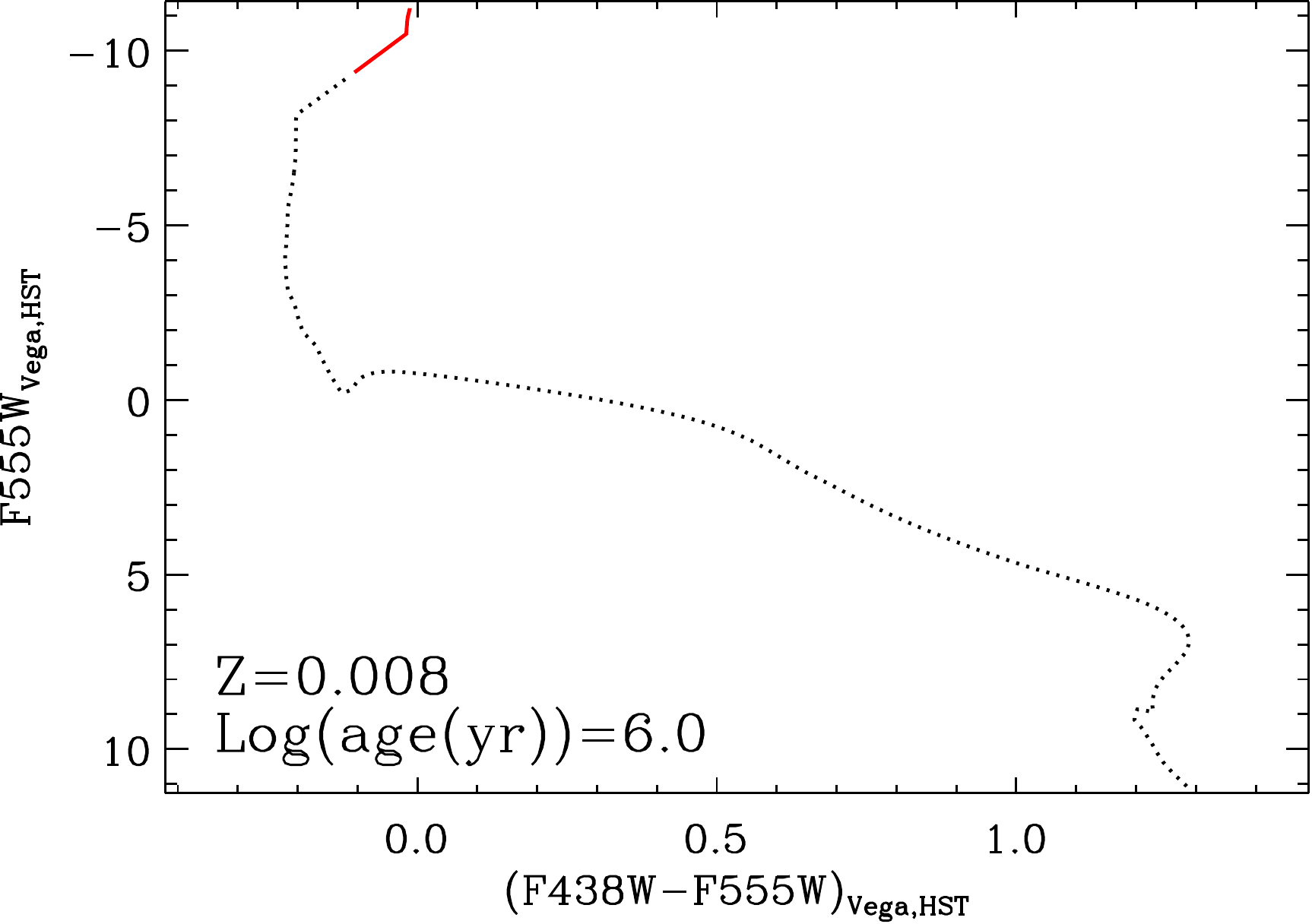}
\vspace{0.2cm}
\includegraphics[scale=0.44,angle=0]{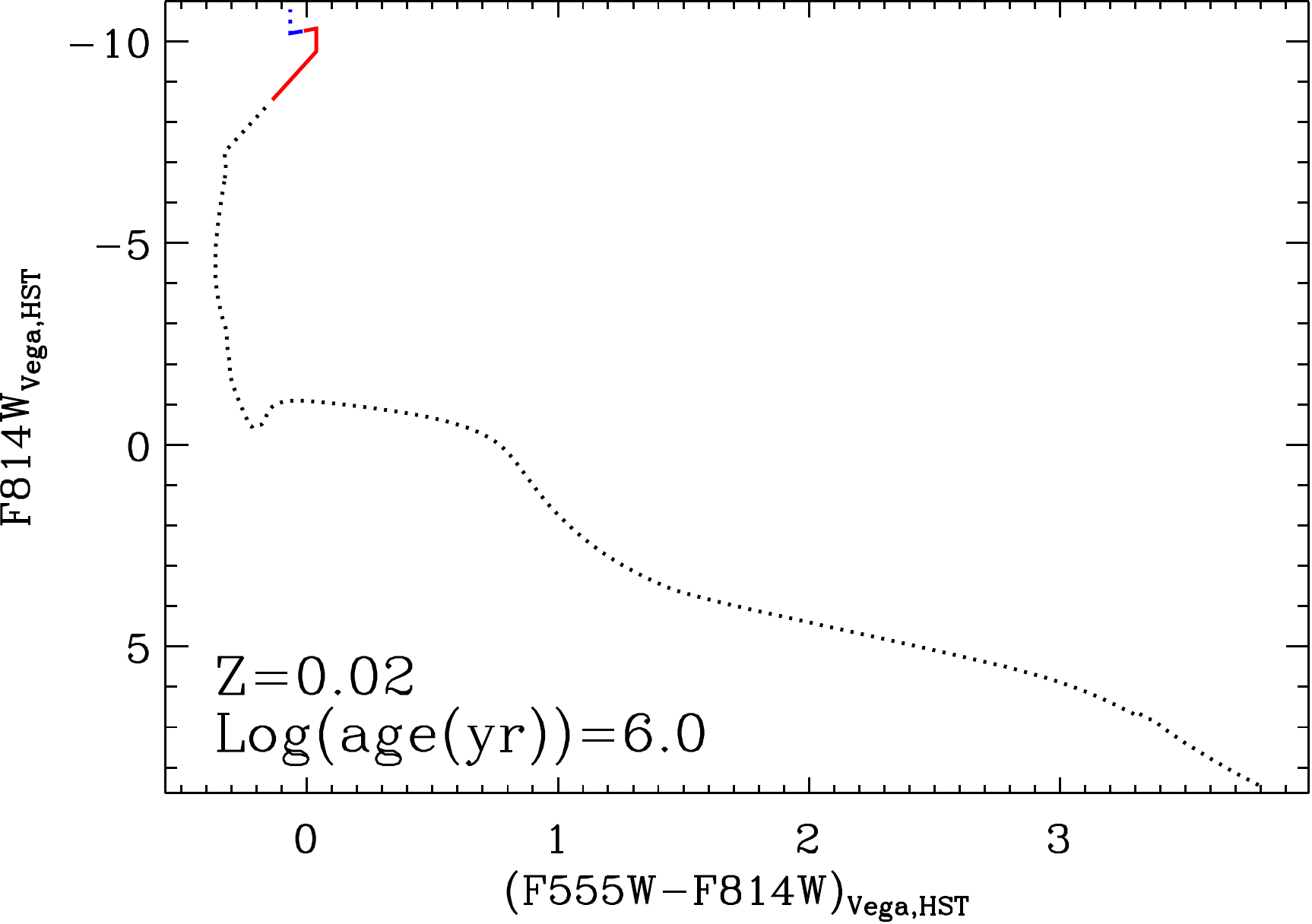}~
\includegraphics[scale=0.44,angle=0]{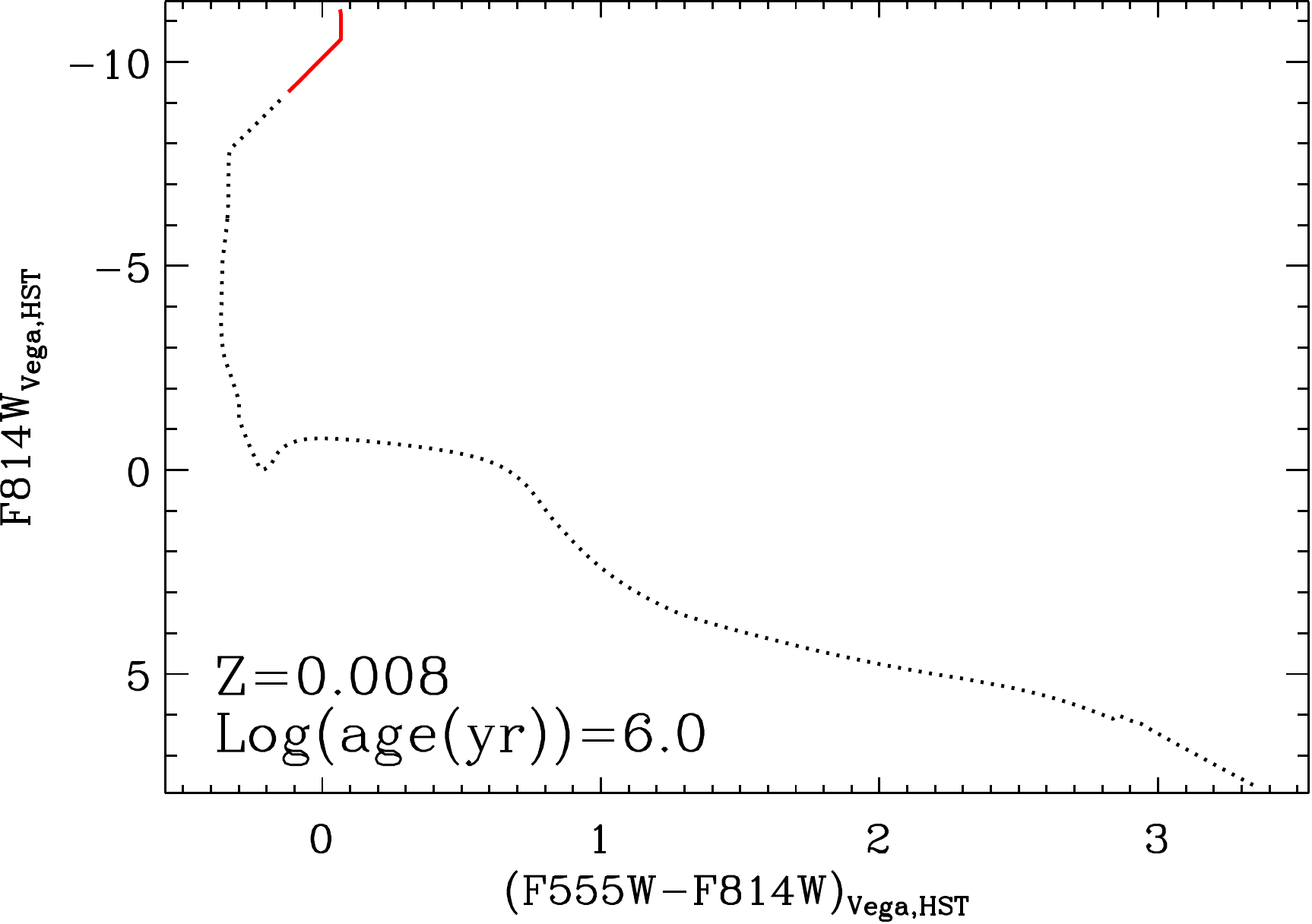}
\vspace{0.2cm}
\includegraphics[scale=0.44,angle=0]{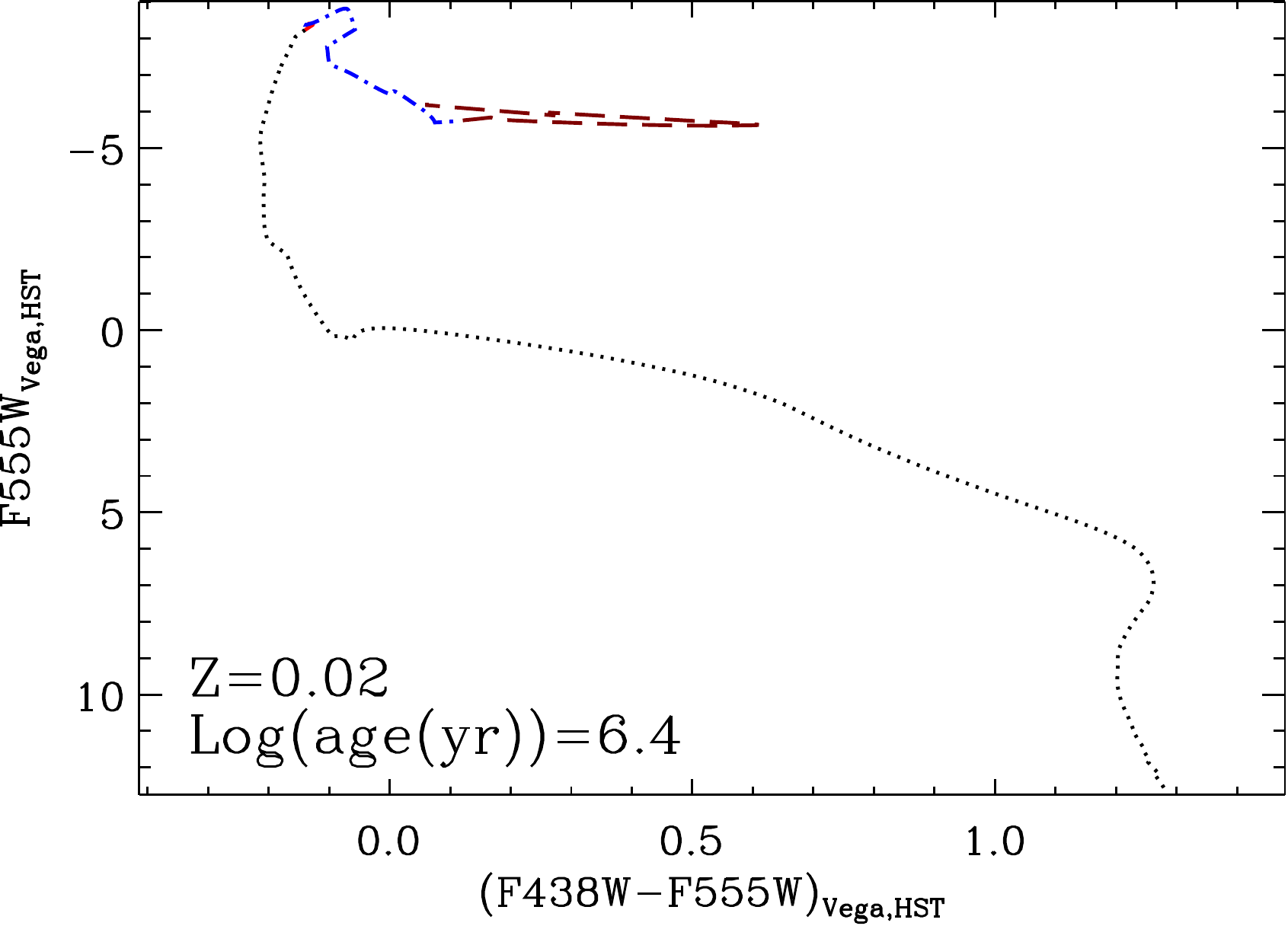}~
\includegraphics[scale=0.44,angle=0]{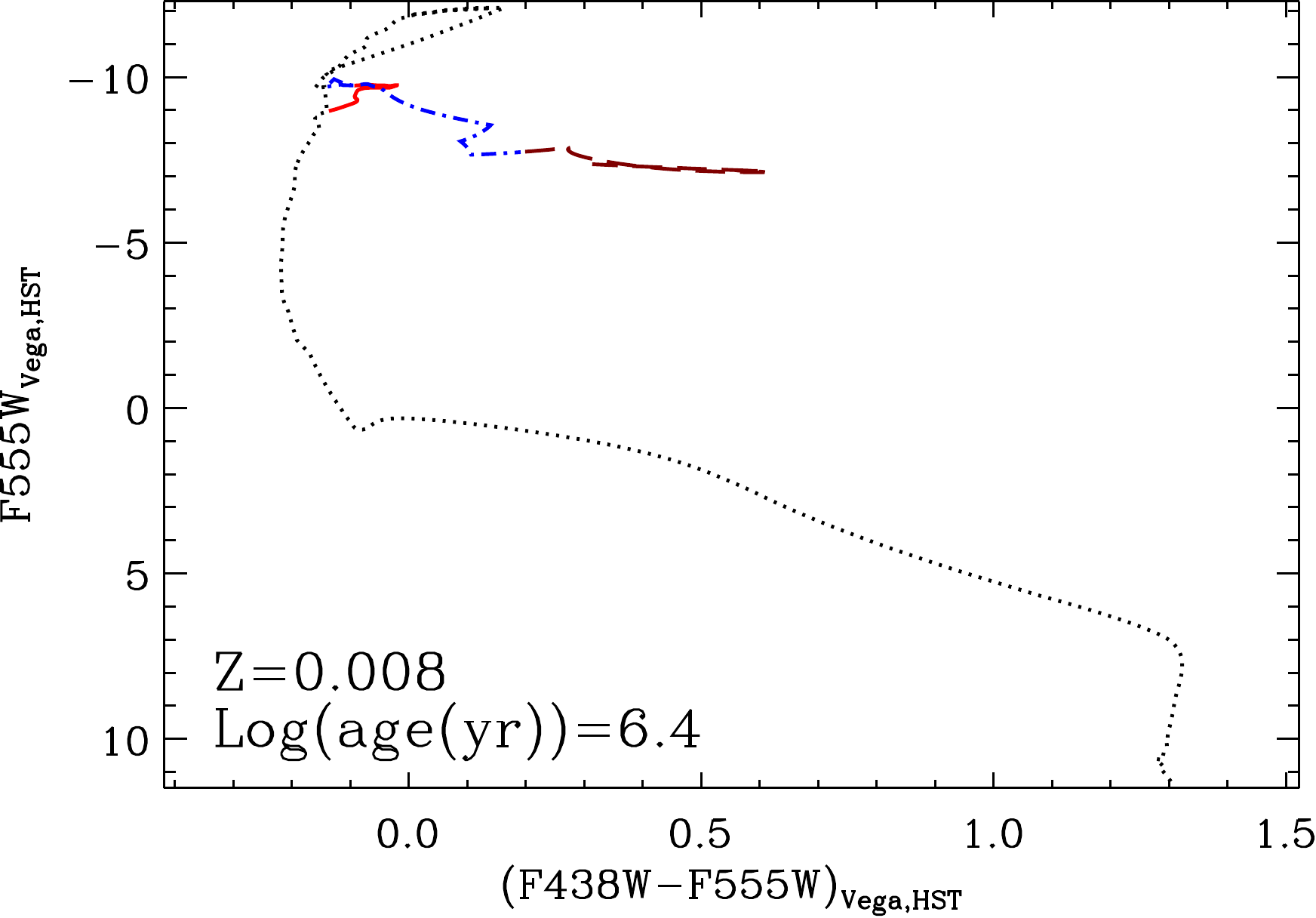}
\vspace{0.2cm}
\includegraphics[scale=0.44,angle=0]{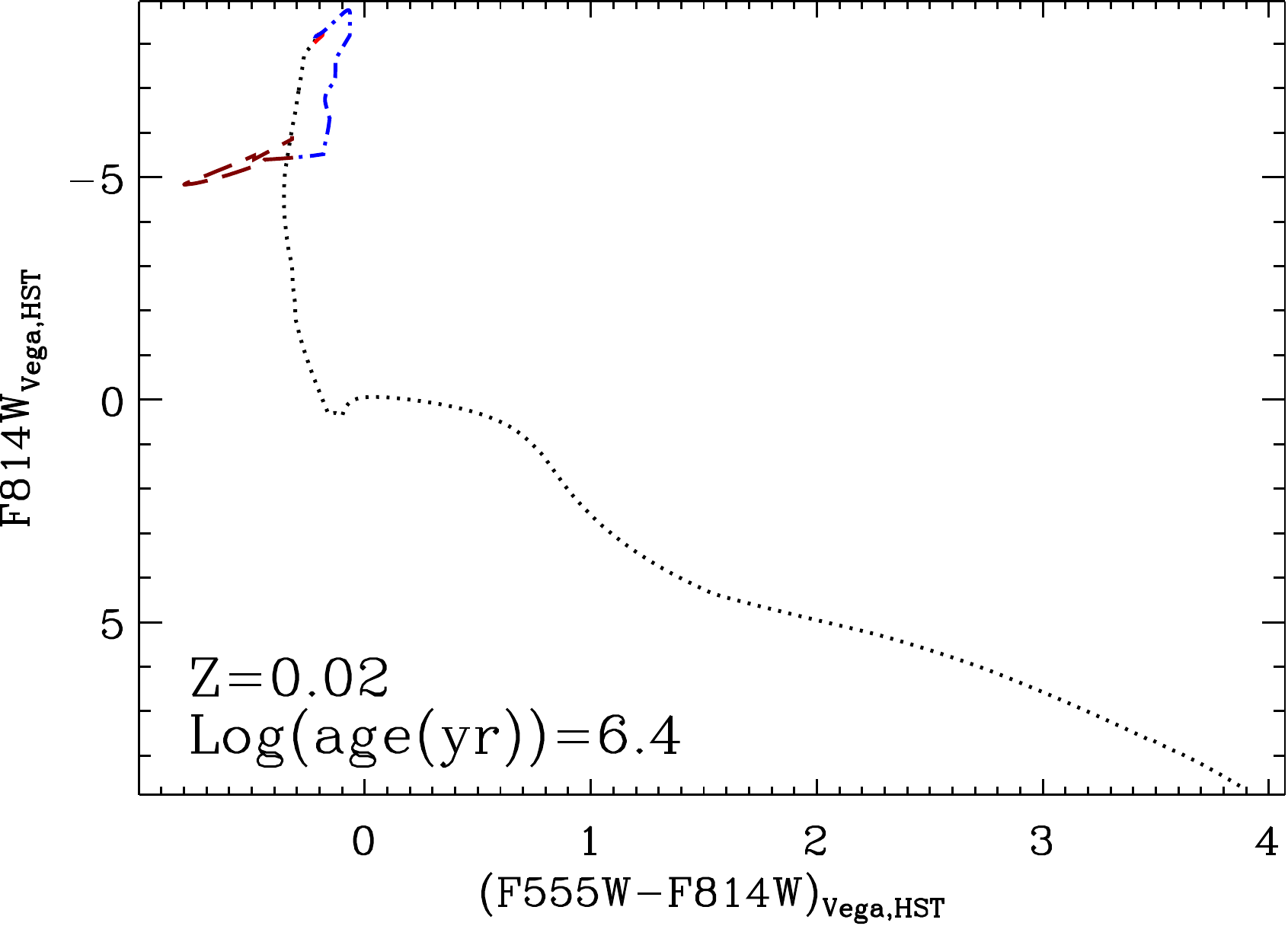}~
\includegraphics[scale=0.44,angle=0]{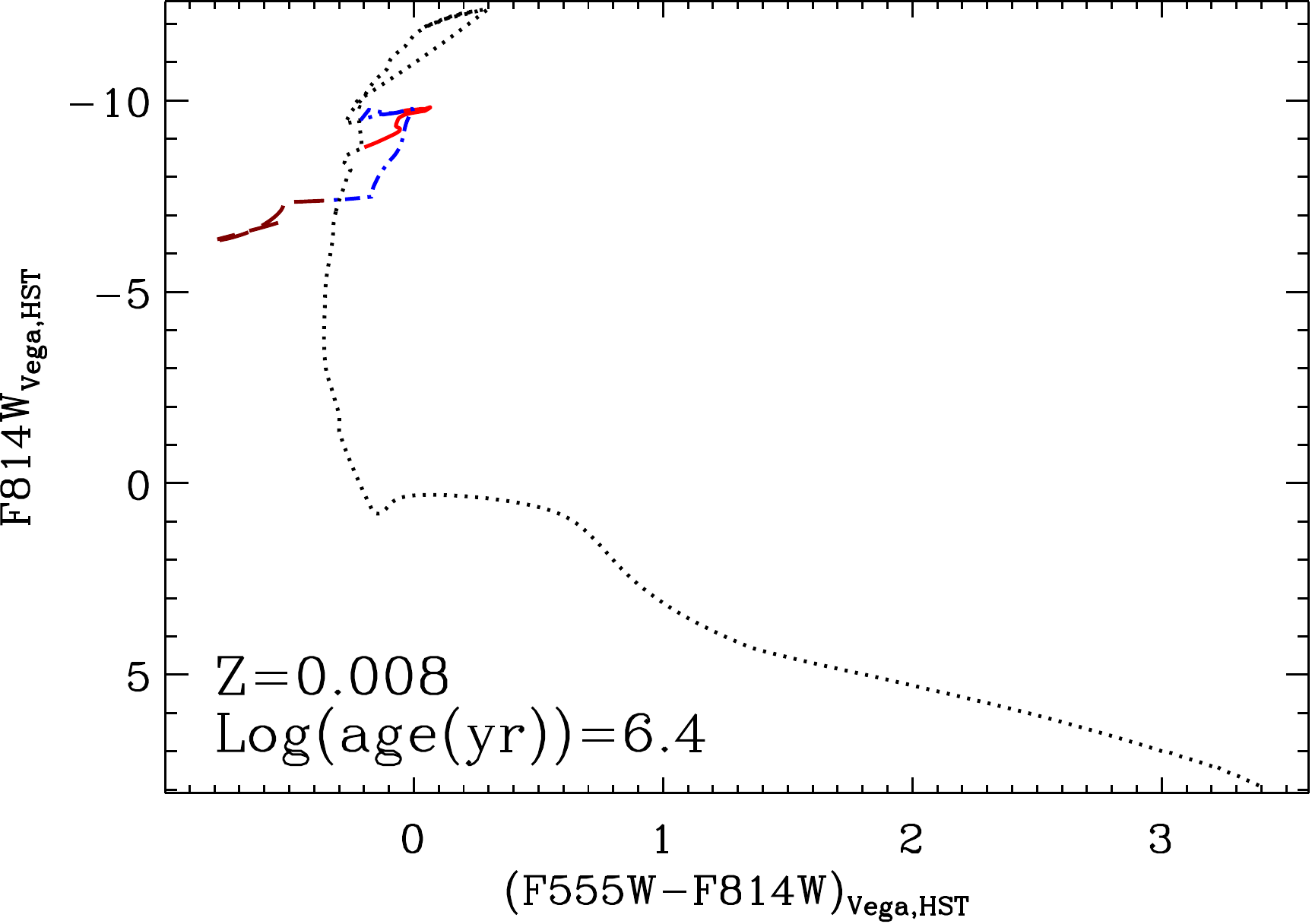}
\caption{Colour-magnitude diagrams in HST/WFC3 broad bands for isochrones of 1\,Myr (upper four panels)
  and 2.5\,Myr (lower four panels) with Z=0.02 (left panels) and Z=0.008 (right panels). The meaning of colours
and line styles  along the isochrones is the same as in
  figure~\ref{massive_tracks_conlib}.
\label{iso_cmd}}
\end{figure*}
\subsection{Cool stars}
As reviewed by \citet{Allard1997}, the atmospheres of cool stars are
dominated by the formation of molecules and eventually at very low
temperatures by dust condensation that can affect the spectral shape
significantly. A suitable set of 1D, static spherical atmosphere
spectral models accounting for the above effects has been developed in
the recent years and is continuously maintained by the Phoenix group
\citep{Phoenix}. Among the different suites of libraries, we use the
BT-Settl models which contain the most updated and complete stellar
parameter grid and are well tested against observations
\citep{Phoenix}.  The full BT-Settl models are provided for
$2600\,$K $\leq \teff < 50000\,K$, $0.5 < \logg < 6$,
and metallicities $0.000003 \lesssim Z \lesssim 0.04$.
We adopt such models for temperatures $\teff \leq 6000\, K$.
 These tables for cool stars have already been used in
\citet{Chen2014}, in the context of low and very low mass stars.

\subsection{Synthetic bolometric correction tables}
A concern for the libraries described above is that the different sets
({\sl WM-basic}, PoWR, ATLAS9 and BT-Settl) are calculated with
different metallicities and different partitions of heavy elements.
In order to obtain homogeneous spectral libraries, at least as far as
the metallicity is concerned, we first calculate the global
metallicity from the detailed abundance values provided by each group.
We note, for example, that the BT-Settl models adopt a solar partition
\citep{AGSS2009} different from the one adopted in PARSEC
\citep{Caffau2011} and that, at low metallicities, their partitions
are $\alpha$~enhanced.  On the contrary, for our {\sl WM-basic}
models, we use the same metallicities as in PARSEC.  We then
interpolate each set of spectra on the global metallicity grid defined
by the PARSEC models.
Since PoWR models are provided only for three
typical metallicities, our Galaxy, the LMC and the SMC, we use Galaxy
models for $Z\geqslant 0.01$, LMC models for $0.01>Z>0.006$ and SMC
models for $Z\leqslant 0.006$.
Unfortunately this procedure may introduce some error because
the partition of heavy elements is not the same in the different libraries.
For example
when we re-scale the \citet{Caffau2011} and the \citet{GS98}
 partitions to a metallicity of Z=0.02, we find a difference
of $\sim$17\% in the Iron content and $\sim$5\% in the Oxygen content.
However we verified that these variations have negligible effects on the broad
band colours of hot spectra.
On the other hand this is the best one can do until a homogeneous set
of model atmospheres (i.e. with the same partition of heavy elements)
is computed on a range of parameters broad enough.
The ranges of effective temperatures encompassed by different
spectral libraries are marked by vertical lines in the HR diagrams
shown in figure~\ref{massive_tracks_conlib}, for $Z=0.02$ (left panel) and
$Z=0.008$ (right panel), respectively.  Overplotted are selected stellar
evolutionary tracks from \Mini=$20\,M_\odot$ to \Mini=$350\,M_\odot$.  The
combined atmosphere models, that share the same global metallicity of
the corresponding evolutionary tracks as shown in the HR diagram, provide
a very good coverage in terms of effective temperatures and gravities.
They allow for an optimal calculation of bolometric correction
(${\rm BC_\lambda}$) tables that are used to convert from theoretical to
observational diagrams. The details of this process are thoroughly
described in \citet{Girardi2002} and \citet{Chen2014}, and are not
repeated here.  While for the evolutionary tracks we only consider the
photospheric magnitudes, the corresponding isochrones also account for
the effect of possible circumstellar dusty envelopes following the
dust calculation recipes described in~\citet{Marigo2008} and ~\citet{Nanni2013,Nanni2014}.

\section{Discussions and Conclusions}

We present new evolutionary tracks of massive stars for a broad range
of metallicities, 0.0001 $\leq~Z~\leq0.04$ and for initial masses up
to \Mini=$350\Msolar$.  At super-solar metallicity, the models extend up
to \Mini=$200\Msolar$ (Z=0.03) and \Mini=$150\Msolar$ (Z=0.04), respectively.
The new models complement the already published PARSEC data base
\citep{PARSEC} and supersede the old Padova evolutionary tracks of
massive stars which are more than 20 years old.  The stellar models
are evolved from the pre-main sequence phase to the central Carbon
ignition.  The mass grid is very well sampled and it is fully adequate
to perform detailed investigations of very young stellar systems both
from the point of view of the resolved populations and from their
integrated properties.  

We revise the scheme adopted for including the mass-loss rate in the
evolution of massive stars, by combining recent recipes found in the
literature. Among the new recipes, particularly important is the
enhancement of the mass-loss rate when the luminosity approaches the
Eddington limit. We show that with the recent formulation by
\citet{Vink2011}, the models naturally reproduce the observed
Humphreys-Davidson limit observed in the Galactic and LMC HR
diagrams. In previous Padova models this limit was used as a threshold
to enhance the mass-loss rates, independently from the metallicity.
In this paper, the role of the metallicity is described by means of a
simple recipe derived from the models presented in
\citet{Grafener2008}.  The metallicity dependence of the mass-loss
rate is now described by a power law with an exponent $\alpha$ which
depends on $\Gamma_e$ (equation~(\ref{gamma})), the ratio between the
stellar luminosity and the Eddington luminosity. When $\Gamma_e$
approaches unity the metallicity dependence drops significantly,
allowing for relatively high mass-loss rates also in stars of low
metallicity.  At lower values of $\Gamma_e$ the usual exponent is
recovered, $\alpha$=0.85.  While the models of \citet{Grafener2008}
refer particularly to the WR phase, it has already been shown that
there is a continuity between the new \citet{Vink2011} formalism and
the \citet{Grafener2008} results for WNL stars.  The result of the new
mass-loss formulation is that the Humphreys-Davidson limit shows a
clear dependence with metallicity and it even disappears at very low
metallicities.

Compared to our previous models with the same
metallicity~\citep{Fagotto_etal94, Bressan1993}, the major difference
concerns the WR phases.  The new models evolve in the late WR stage, at
$\logteff > 5.0$, with luminosities which are significantly higher than
those of the old Padova tracks ($\sim 0.9\,{\rm dex}$ in $\logL$).
These differences result from the different mass-loss recipes used in
the advanced WR phases.  Since the lifetimes of the models in this
phase remain more or less the same, the new models should contribute
much more to the hard ionizing photons.  At lower metallicity,
Z$\leq$0.008, the new mass-loss formulation introduces significant
differences even in the earlier phases. Our models compare well with
the recent evolutionary tracks of non-rotating stars of similar
metallicity, computed with FRANEC~\citep{Chieffi_etal2013}.

Besides evolutionary tracks, we provide new tables of bolometric
corrections that allow for the conversion from the theoretical HR to the
observed colour-magnitude diagrams.  For this purpose, we assemble existing
atmosphere libraries, ATLAS9, PoWR, Phoenix and new atmosphere models
calculated on purpose with the {\sl WM-basic} code.  We merge the
different libraries with interpolation on a global metallicity scale,
providing quite homogeneous tables of bolometric corrections, at
several metallicities.

An example of the colour-magnitude diagrams of the tracks of \Mini=$100M_\odot$ and
\Mini=$50M_\odot$ is shown in figure~\ref{cmd_track}, for Z=0.02 (left
panels) and Z=0.008 (right panels), respectively.  In the figure, we
highlight the different evolutionary phases with different colours and line styles.
Black dotted lines indicate the evolution precedent of the WR stages, while
the red solid lines are used to indicate the transition phase from the LBV
phase to the late WR stars (WNL-H50 for $Z=0.02$ or WNL-H40 for
$Z=0.008$ in the PoWR notation).  The blue dash-dotted lines are used for the
other WN stages and the brown dashed lines are for the final WC/WO stages.

As can be seen in the figure, the optical colours show some evident
jumps when the star type changes from WN to WC. This is likely to be
caused by the appearance of different strong emission lines in
different sub-types.  This can be seen from figure~\ref{compare_WC_WN},
where we compare the spectra of one WC star (black solid line) and one WN (red dashed line)
star with the same effective temperature. We show the optical region
sampled by the broadband filters F438W, F555W, and F814W of the
HST/WFC3 system.  To better clarify this point we have used in the
comparison the high resolution CMFGEN spectra~\citep{CMFGEN}, which
better show how the emission lines affect the spectra and the
broadband colours.  Both spectra show a strong emission HeII(4700\AA)
line, falling within the F555W passband and touching the border of
the F438W passband.  The differences between the WN and the WC
spectral types are mainly on the contribution of the strong
CIV(5800\AA) line within the F555W passband, in the latter type.
Thus, the flux in the F555W passband is heavily enhanced in the case
of the WC star with respect to that of the WN star, and
correspondingly the sudden transition from the WN to the WC types is
accompanied by a jump in the F438W-F555W and F555W-F814W colours.  It
is important to stress that the CIV(5800\AA) doublet is very sensitive
to the adopted atmosphere parameters which somehow challenges its
predictability (private communication with Helge Todt, Wolf-Rainer Hamann
 and G{\"o}tz {Gr{\"a}fener}).  Furthermore, we have to
note that the transition between WN and WC spectral types is
smoother in the tracks than in the spectra, because the latter are
computed only at discrete values of the elemental abundances, and this
could enhance the effect, at least in terms of evolutionary speed.

Using the evolutionary tracks we compute new isochrones of young
stellar populations, with the same procedure already described
in~\cite{PARSEC}.  A few examples are shown in the theoretical HR
diagram of figure~\ref{isochroneZ}, from very young to very old ages
and for Z=0.02 (left panel) and Z=0.008 (right panel), respectively.  By means of
the new bolometric correction tables, we convert theoretical
isochrones into observational magnitudes/colours, in the same way used
for the evolutionary tracks.  In figure~\ref{iso_cmd} we show the
colour-magnitude diagrams of isochrones at very young ages, 1\,Myr and
2.5\,Myr, and for Z=0.02 and Z=0.008.  The colour codings are the same
as in figure~\ref{cmd_track}, but for the pre-main sequence which is
drawn in light gray.

A preliminary comparison of the new models with the colour-magnitude
diagrams of star-forming regions in nearby low metallicity dwarf
irregular galaxies, has already been performed
in~\citet{Tang_etal2014}.  The new models of massive stars will be
used to compute the integrated properties of young star forming
regions (Obi et al. in preparation).  The full sets of evolutionary
tracks can be downloaded from
\url{http://people.sissa.it/~sbressan/parsec.html}.  The isochrones
can be downloaded from \url{http://stev.oapd.inaf.it/cgi-bin/cmd}.

\section*{Acknowledgments}
We thank the anonymous referee for the suggestions
which helped to improve the paper.
We thank A.W.A. Pauldrach for the help on running {\sl WM-basic},
D.J. Hillier for the help with the CMFGEN code,
M. Limongi for providing their tracks.
We specially thank H. Todt, W.-R. Hamann, and G. Gr{\"a}fener
for their extensive help on the PoWR data base.
We thank S. Charlot, X.T. Fu and Z.Y. Zhang for helpful discussions.
Y. Chen and A. Bressan, and L. Girardi acknowledge the financial support
from INAF through grants PRIN-INAF-2014-14.
A. Bressan, L. Girardi and P. Marigo
acknowledge the support from the ERC Consolidator Grant funding scheme
({\em project STARKEY}, G.A. n. 615604).
P. Marigo acknowledges
the support from the University of Padova, ({\em Progetto di Ateneo 2012},
ID: CPDA125588/12).
X. Kong and Y. Chen acknowledge financial supports
from the National Natural Science Foundation of China (NSFC,
Nos.11225315, 1320101002, 11433005, and 11421303), the Strategic
Priority Research Program ``The Emergence of Cosmological Structures''
of the Chinese Academy of Sciences (No. XDB09000000), the Specialized
Research Fund for the Doctoral Program of Higher Education (SRFDP,
No. 20123402110037), and the Chinese National 973 Fundamental Science
Programs (973 program) (2015CB857004).

This research has made use of NASA's Astrophysics Data System Bibliographic Services.

\bibliographystyle{mn2e_new}
\bibliography{reference}

\label{lastpage}

\end{document}